\documentclass[aps,pra,twocolumn,superscriptaddress,showpacs]{revtex4-1}
\usepackage{graphicx,bm,color,float,textcomp,mathtools,times}
\newcommand\sho{$\rm SrHo_2O_4$}
\newcommand\sgo{$\rm SrGd_2O_4$}
\newcommand\seo{$\rm SrEr_2O_4$}
\newcommand\sdo{$\rm SrDy_2O_4$}
\newcommand\sybo{$\rm SrYb_2O_4$}
\newcommand\sRo{SrRE$_2$O$_4$}
\newcommand\afm{antiferromagnetic}
\newcommand\KP{${\bf k}^\prime \! = \! [0 \frac{1}{3} \frac{1}{3}]$}
\newcommand\KO{${\bf k} = 0$}

\begin{document}
\title{Evolution of spin correlations in $\mathbf{SrDy_2O_4}$ in an applied magnetic field}
\author{O.~A.~Petrenko} \affiliation{Department of Physics, University of Warwick, Coventry CV4 7AL, United Kingdom}
\author{O.~Young} \affiliation{High Field Magnet Laboratory, Institute for Molecules and Materials, Radboud University, 6525 ED Nijmegen, The Netherlands}
\author{D.~Brunt} \affiliation{Department of Physics, University of Warwick, Coventry CV4 7AL, United Kingdom}
\author{G.~Balakrishnan} \affiliation{Department of Physics, University of Warwick, Coventry CV4 7AL, United Kingdom}
\author{P.~Manuel} \affiliation{ISIS Facility, STFC Rutherford Appleton Laboratory, Chilton, Didcot, OX11 0QX, United Kingdom}
\author{D.D.~Khalyavin} \affiliation{ISIS Facility, STFC Rutherford Appleton Laboratory, Chilton, Didcot, OX11 0QX, United Kingdom}
\author{C.~Ritter} \affiliation{Institut Laue-Langevin, Boite Postale 156X, F-38042 Grenoble Cedex 9, France}
\begin{abstract} 
The development of short- and long-range magnetic order induced in a frustrated zig-zag ladder compound \sdo\ by an applied field is studied using neutron diffraction techniques.
In zero field, \sdo\ lacks long-range magnetic order down to temperatures as low as 60~mK, and the observed powder neutron diffraction (PND) patterns are dominated by very broad diffuse scattering peaks.
Single crystal neutron diffraction reveals that the zero-field magnetic structure consists of a collection of \afm\ chains running along the $c$ axis and that there is very little correlation between the chains in the $ab$ plane.  
In an applied magnetic field, the broad diffuse scattering features in PND are gradually replaced by much sharper peaks, however,  the pattern remains rather complex, reflecting the highly anisotropic nature of \sdo. 
Single crystal neutron diffraction shows that a moderate field applied along the $b$ axis induces an {\it up-up-down} magnetic order associated with a 1/3-magnetisation plateau, in which magnetic correlation length in the $ab$ plane is significantly increased, but it nevertheless remains finite.
The resolution limited ${\bf k}=0$ peaks associated with a ferromagnetic arrangement appear in powder and single crystal neutron diffraction patterns in fields of 2.5~T and above.
\end{abstract}
\maketitle
\section{Introduction}
\begin{figure}[tb] 
\centering
\vspace{-3mm}
\includegraphics[width=\columnwidth]{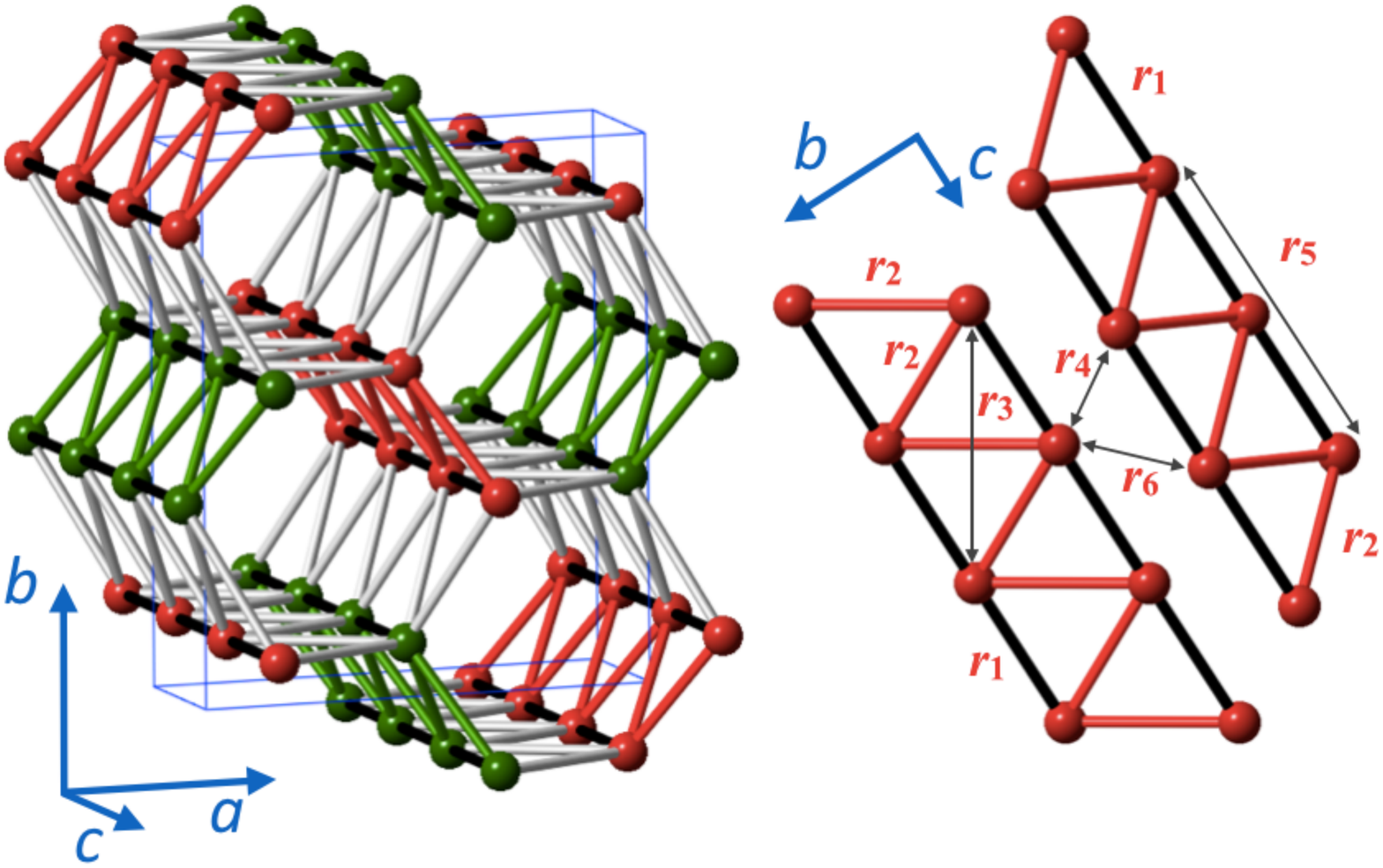}
\vspace{-3mm}
\caption{(Colour online) Positions of the magnetic Dy$^{3+}$ ions within \sdo.
		Left panel depicts the honeycomb-like arrangement of the magnetic ions with two crystallographically inequivalent Dy$^{3+}$ sites shown in green (Dy1) and red (Dy2).
		The box represents a crystallographic unit cell of the $Pnam$ space group.
		Right panel emphasises the zigzag ladders formed by ions on the Dy2 site when viewed along the $a$~axis and shows some of the interionic distances labelled as $r_i$.}
\label{Fig1_structure}
\end{figure}

In highly frustrated magnetic systems, the degeneracy of the ground state (due to the effective cancelation of the main exchange couplings) makes such systems susceptible to a variety of much weaker interactions~\cite{Diep_2005,Buschow_2001, Lacroix_2011,Ramirez_1994,Greedan_2001}.
These interactions (typically ignored in systems without appreciable frustration) often result in the formation of unusual magnetically ordered phases, both with short-range and long-range correlations.
The family of rare-earth (RE) strontium oxides, \sRo, provides an excellent opportunity to study the interplay between the geometrical frustration of the exchange interactions, the dipolar coupling and the single-ion magnetic properties of different RE elements~\cite{Karunadasa_2005}.
The resulting magnetic ground states are often very complex~\cite{Petrenko_2014} and consist of two different components corresponding to the two inequivalent positions of the magnetic RE ions in the unit cell (see Fig.~\ref{Fig1_structure} which depicts the positions of the magnetic Dy$^{3+}$ ions within \sdo ).

The local environment is different for the two RE sites in all the \sRo\ compounds, therefore the magnetic moments on them, which are often Ising in nature, have different easy-axis directions. 
In \seo, a ${\bf k}=0$ long-range \afm\ order with magnetic moments parallel to the $c$ axis on one Er site and a quasi one-dimensional (1D) short-range \afm\ order with the moments along the $a$ axis on another Er site are found to coexist down to the lowest temperatures~\cite{Petrenko_2008,Hayes_2011}.
In \sho, the magnetic structure is similar to that of \seo, but the ${\bf k}=0$ component has only a limited correlation range and the magnetic moments in the quasi-1D structure point along the $b$ axis~\cite{Young_2012,Young_2013}.
In \sybo, the structure is reported to be a noncollinear ${\bf k}=0$ antiferromagnet in which the magnetic moments of two inequivalent Yb ions lie in the $ab$ plane, but have different moment sizes and directions~\cite{Quintero_2012}.
The magnetic structure of \sgo\ at different temperatures is yet to be determined.
This is the only compound in the family with a well-defined double phase transition -- bulk property measurements give clear indications of magnetic ordering at 2.73~K and a further transition at 0.48~K~\cite{Young_2014}.

In sharp contrast to the family members discussed above, \sdo\ does not show any signs of a magnetic phase transition down to the lowest experimentally available temperatures.
In zero field, heat capacity measurements indicate that this compound appears to be magnetically disordered down to at least 0.39~K~\cite{Cheffings_2013}, while powder neutron diffraction (PND) reveals only diffuse scattering peaks even at 50~mK~\cite{Poole_2014}.
However, heat capacity~\cite{Cheffings_2013} and magnetisation~\cite{Hayes_2012} measurements both indicate a highly anisotropic behaviour in an applied magnetic field, including a sequence of magnetically ordered phases for a field parallel to the $b$ axis.
Recent ultrasound investigations mapped out the high resolution $H-T$ phase diagrams for the magnetic field applied along the $b$ and $c$ axes~\cite{Bidaud_2016} and suggested that for $H\parallel b$ the long-range-ordered magnetic state is contained within a ``dome" extending up to 2.2-2.5~T at lowest temperatures and up to 1.5~K in a field of 1~T.

In this paper we report the investigations on the nature of the field-induced magnetic ordering in \sdo\ using neutron diffraction measurements performed on both powder and single crystal samples.
Our main results can be grouped into three categories: (i) zero field, where there is no long-range magnetic order, (ii) intermediate fields (up to and including 2~T) where a ``ferrimagnetic'' state is found,  (iii) higher field (2.5 and 3~T), where a  polarised ``ferromagnetic" state is supported.
Zero field PND data are analysed using the SPINVERT program~\cite{Paddison_2013}, which implements a reverse Monte Carlo (RMC) algorithm.
RMC analysis reveals the patterns of spin-spin correlations in \sdo\ and allows for a more detailed comparison with the magnetic structures of other \sRo\ members.

For $H \parallel b$, an evolution of magnetic structure from collections of weakly-correlated \afm\ chains, through a much more correlated {\it up-up-down} phase corresponding to a magnetisation plateau in intermediate fields, to an effective ferromagnetic state in higher fields is clearly observed in single crystal diffraction experiments.
Remarkably, the {\it up-up-down} phase with the magnetic moments parallel to $b$ axis is also clearly seen through the Rietveld refinement of the PND patterns in intermediate magnetic fields despite the inevitable presence of other magnetic phases in a polycrystalline sample exposed to an external magnetic field.
We also find that the higher-field ferromagnetic state is not a fully polarised phase, as only magnetic moments on one Dy site participate in the formation of the field-induced magnetic order, while the moments on the other site remain largely disordered.  

These findings should be helpful for the eventual development of realistic models of the magnetic interactions in the \sRo\ family of compounds.
Some estimates of the interaction ratios in \sdo\ and \sho\ were made from the crystal-field levels~\cite{Poole_2014}.
Further, the anisotropic next-nearest-neighboor model was applied to \sho\ and the relevant parameters estimated~\cite{Wen_2015}.
For \seo, the latest estimates of the exchange and dipolar interactions were derived from the four-particle self-consistent model~\cite{Malkin_2015}.
Despite all the above and the recent theoretical developments~\cite{Dublenych_2016}, it is fair to state that a full understanding of which interactions are dominant in each compound is still largely missing.

\section{Methods}
\label{sec_methods}
A polycrystalline sample of \sdo\ was prepared from the high-purity starting materials SrCO$_3$ and Dy$_2$O$_3$ following previous reports~\cite{Karunadasa_2005,Petrenko_2008}.
Dysprosium enriched with the $^{162}$Dy isotope (95\%) was used for the powder sample preparations in order to reduce the high neutron absorption associated with the presence of $^{161}$Dy and $^{164}$Dy isotopes in naturally occurring dysprosium.
The single crystal of \sdo\ was grown by the floating zone technique using an infrared image furnace as reported previously~\cite{Balakrishnan_2009}.
Naturally occurring dysprosium was used for the single crystal growth.
Smaller crystals from the same growth were used in preceding magnetisation~\cite{Hayes_2012} and heat capacity~\cite{Cheffings_2013} investigations. 

High-resolution PND experiments were conducted using a wavelength of 1.594~\AA\ on the D2B diffractometer at the Institut Laue-Langevin, France.
Diffraction patterns were collected in zero field at various temperatures between 60~mK and 10~K as well as in an applied field of up to 2.5~tesla at a base temperature of 60~mK.
A copper sample holder was used in order to provide better heat exchange at low temperatures.
Additional high resolution room-temperature measurements were performed on the D1A diffractometer to refine the scattering length of the Dy-isotope.

Single crystal neutron-diffraction measurements were made on the WISH diffractometer~\cite{WISH} at the ISIS facility at the Rutherford Appleton Laboratory (United Kingdom) in a range of temperatures from 60~mK to 10~K and in fields of up to 3~T.
The sample (0.44~g) was fixed to an oxygen free copper holder with the $b$ axis vertical defining the horizontal $(h0l)$ scattering plane.
With a continuous array of position sensitive detectors available on WISH, substantial coverage of the out-of-plane scattering, essential for the results reported in this paper, was also possible, as we were able to reach the reflections between and including the $(h1l)$ and $(h\bar{1}l)$ planes.
Two sample positions, different by 35 degrees in rotation around the vertical axis, were used in order to optimise the neutron flux around the $(0 0 \frac{1}{2})$ and $(2 0 0)$ positions.

For the experiments on both D2B and WISH diffractometers, a dilution refrigerator inside a vertical-field cryomagnet provided the necessary sample environment. 

The \textsc{fullprof} suite of programs (including \textsc{k-search} and \textsc{basireps}) was used for refinements of the magnetic phases~\cite{FULLPROF}.
The \textsc{spinvert} program~\cite{Paddison_2012,Paddison_2013} was used to refine the zero-field magnetic diffuse scattering patterns.
Here, the crystallographic unit cell is used to generate a supercell ($15 \times 15 \times 40$, total number of spins 72000), and at first a random Ising spin is assigned to each Dy site (with moments on one site parallel to the $b$~axis, and parallel to the $c$~axis on the other) from which the sum 
\begin{equation}
\chi^2 = W \sum_Q \bigg[ \frac{s I_{\rm calc}(Q) - I_{\rm exp}(Q)}{\sigma(Q)}\bigg]^2 
\end{equation}
is calculated.  Here, $I(Q)$ is the 
magnetic scattering intensity, $\sigma(Q)$ is the experimental uncertainty, $W$ is an empirical weighting factor, and $s$ is the scale factor.
An RMC algorithm is then used to fit the data by minimising the sum of the squared residuals -- {\it i.e.} by randomly flipping a spin and then re-calculating $\chi^2$ until no more variation is observed.
Generally, 100 moves were proposed per spin, and these moves were accepted or rejected based on the Metropolis algorithm until the best fit to the data is obtained.
The value of the intensity scale factor, $s$, was refined when fitting the base temperature data, and this was kept fixed when fitting the data for higher temperatures. 
Changing around the direction of the spins on the two crystallographically inequivalent sites ({\it i.e.} with moments on one site now parallel to the $c$~axis, and parallel to the $b$~axis on the other) makes no quantitative difference to $\chi^2$, and qualitatively the simulated single crystal diffraction patterns look identical.
The scattering patterns and spin correlations were calculated using the programs \textsc{spindiff} and \textsc{spincorrel}~\cite{Paddison_2012, Paddison_2013}.
To minimise statistical noise, calculated quantities were averaged over 10 independent spin configurations for each temperature.

\section{Powder neutron diffraction}
\subsection{Zero field PND} \label{Zero_field_PND}
\begin{figure}[tb] 
\centering
\includegraphics[width=\columnwidth]{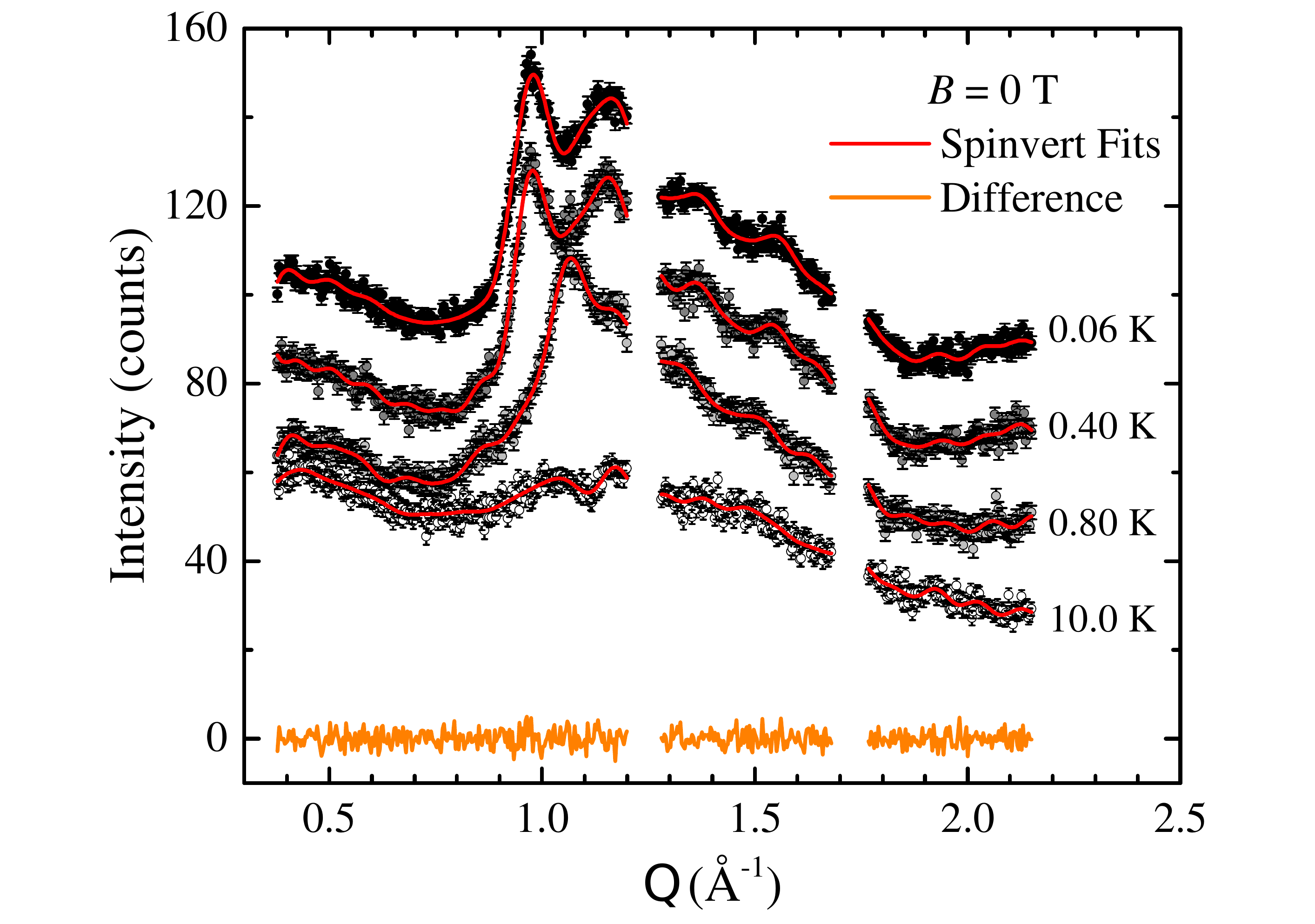}
\vspace{-3 mm}
\caption{(Colour online) Lower-$Q$ part of the PND patterns of \sdo\ at $T=$ 0.06, 0.4, 0.8 and 10~K in zero field.
		The base temperature $T=0.06$~K dataset is on the original experimental intensity scale, the other datasets are consecutively offset by $-20$ units for clarity.
		The difference plot included is for the base temperature of 0.06~K.}
\label{Fig2_H=0T_Tdep}
\end{figure}

PND patterns collected at $T=0.06$, 0.4, 0.8 and 10~K in zero field are displayed in Fig.~\ref{Fig2_H=0T_Tdep}.
The data show very broad diffuse scattering features indicating that there are significant short range correlations between the magnetic moments but no static magnetic order.
At the lowest temperature, the magnetic scattering is dominated by a broad peak around $Q = (00\frac{1}{2})$, which decreases in intensity and shifts to higher $Q$ with increasing temperature. 
By warming up to 10~K, there are no distinct features in the scattering, however, there is still a difference between the obtained data and the form factor expected for purely paramagnetic scattering from Dy.
Such broad features in reciprocal space cannot be fitted using conventional means such as Rietveld refinement, but as their symmetry is governed by the lattice and their modulation depends on the nature of the magnetic interactions, we have used the RMC method described in Section~\ref{sec_methods} to investigate the short-range correlations in \sdo. 
The RMC method (like Rietveld refinements) does not give a unique magnetic structure ``solution", but suggests a model that is compatible with the data.
The $Q$-range used for the RMC is 0.378 to 2.150~\AA$^{-1}$, and two small nuclear Bragg peaks have been omitted from the refinements for all temperatures.
The PND data and RMC fits, displayed in Fig.~\ref{Fig2_H=0T_Tdep}, show very good agreement, with $R_{\rm WP}$ = 1.69, 1.77, 1.71, 1.59\% for 0.06, 0.4, 0.8 and 10~K respectively.
Such a good agreement further justifies the choice of Ising model for the moments on both Dy sites.

\begin{figure}[tb] 
\centering
\vspace{-11 mm}
\includegraphics[width=\columnwidth]{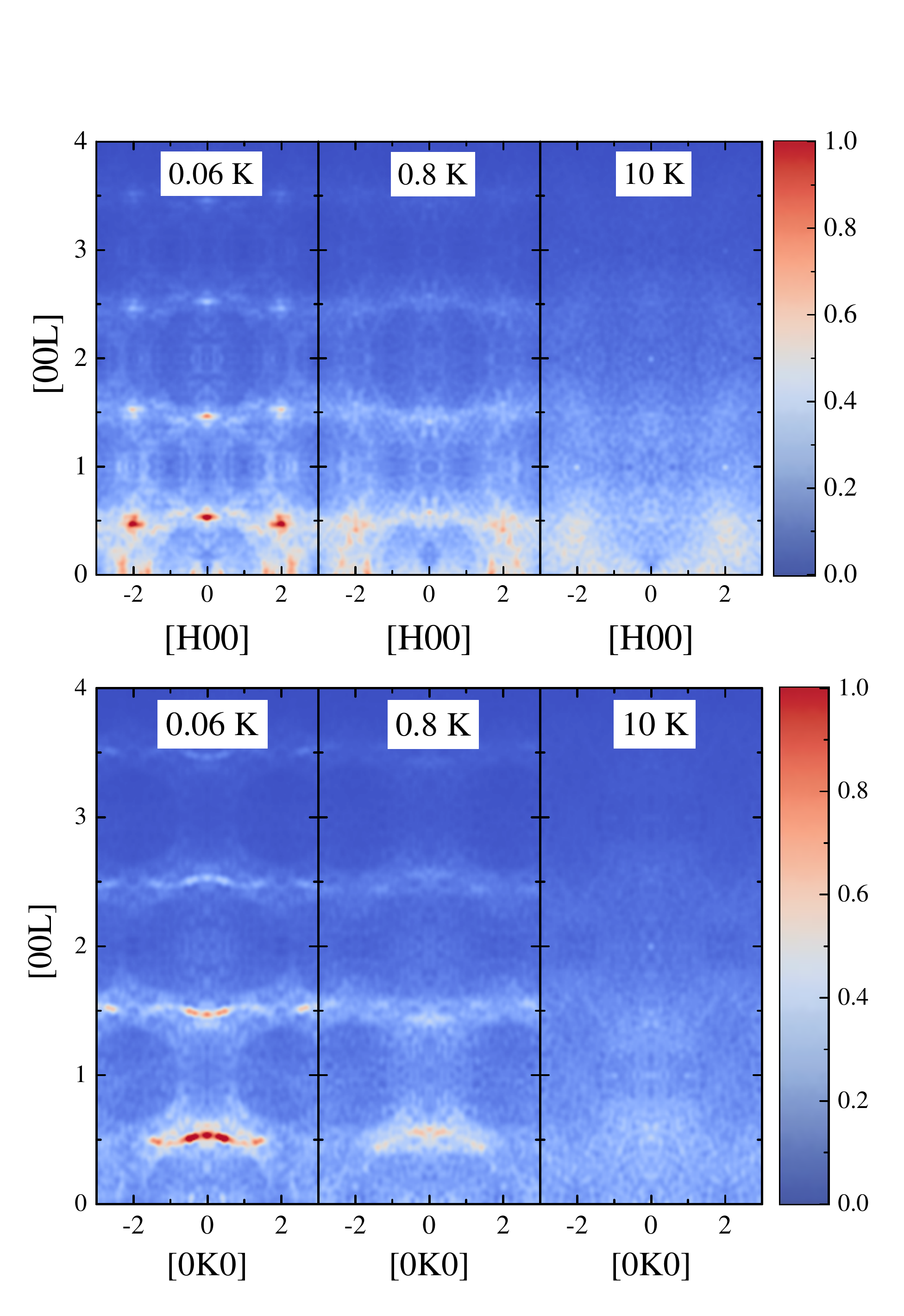}
\vspace{-5 mm}
\caption{(Colour online) The single crystal magnetic diffuse scattering patterns calculated from the refined RMC model in the $(h0l)$, top panel, and $(0kl)$, bottom panel, scattering planes at different temperatures.
			The same intensity scale is used in all the panels.}
\label{Fig3_RMC}
\end{figure}

Calculations of the single crystal scattering patterns from the refined RMC model are shown in Fig.~\ref{Fig3_RMC}.
The magnetic intensity is most pronounced around the $l=\pm \frac{1}{2}$ and symmetry related positions when viewed in both the $(h0l)$ and $(0kl)$ scattering planes.
At low temperatures, the correlation length along the $c$~axis is much longer than along both the $a$ and $b$~axes, and the scattering intensity is heavily modulated along both $h$ and $k$.
The temperature evolution of the scattering shows that there is very little difference between PND patterns taken at 0.06 and 0.4~K (see Fig.~\ref{Fig2_H=0T_Tdep}).
Consequently the calculated single crystal scattering patterns are virtually identical for $T=0.06$ and 0.4~K, therefore the $T=0.4$~K data are excluded from Fig.~\ref{Fig3_RMC}.
On further temperature increase to 0.8~K the lowest-$Q$ peak shifts to a higher scattering vector value and the others become less intense; the associated single crystal scattering profile also becomes less intense and more indistinct, but the higher intensity around the $(h0\frac{1}{2})$, $(0k\frac{1}{2})$ and symmetry related positions remains clearly visible.
Finally by 10~K almost all structure in the scattering profile disappears.
For the ($hk0$) scattering plane (not shown), the magnetic scattering is a lot more diffuse and thus the intensity is more smeared out even at the lowest temperatures. 
These observations suggest that the spin correlations for the Dy1 site are systematically weaker compared to the Dy2 site at each temperature, such that on cooling the magnetic moments parallel to the $b$~axis (Dy2 sites) develop significant correlations first and the moments parallel to the $c$~axis (Dy1 site) require lower temperatures to establish considerable correlations.
Therefore the ${\bf k}=0$ structure on the Dy1 sites in \sdo\ is effectively developing in the presence of local fields caused by the highly-correlated moments on the Dy2 sites.

\begin{figure}[tb] 
\centering
\vspace{-1 mm}
\includegraphics[width=\columnwidth]{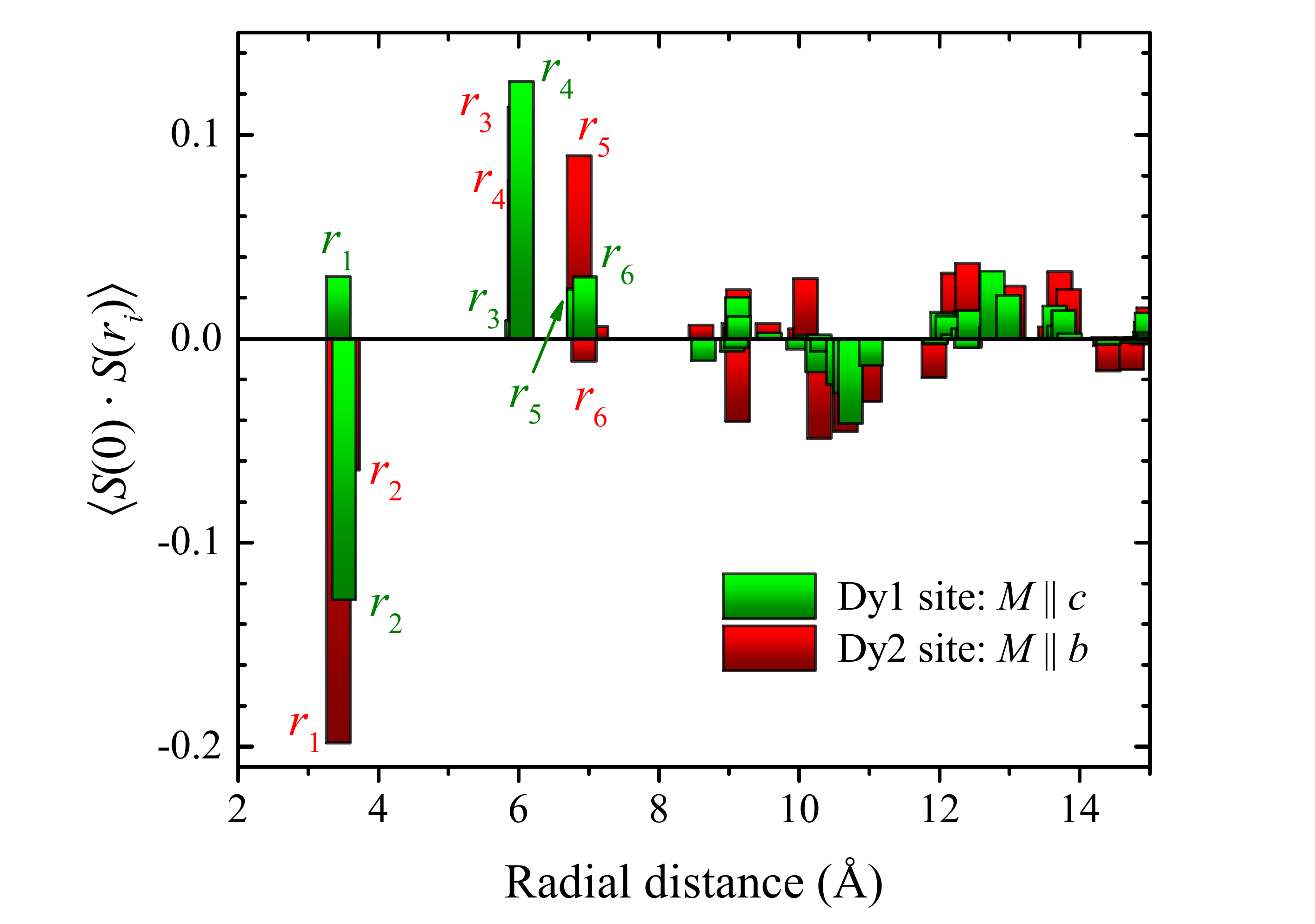}
\vspace{-5mm}
\caption{(Colour online) Spin correlation $\langle S(0) \cdot S(r_i) \rangle$ as a function of the radial distance $r_i$ calculated from the refined RMC model for $T=0.06$~K.
			The index $i$ in $r_i$ refers to the set of the $i^\mathrm{th}$ neighbours (some of them, up to $i=6$, are labeled on the figure).
			The correlation function is split into two groups for spins pointing along the $b$~axis (Dy2 site) and along the $c$~axis (Dy1 site).}	
\label{Fig4_correls}
\end{figure}

The calculated maxima in the scattering intensity are a direct result of the Ising-like nature of the spins and reflect the fact that \sdo\ remains in a disordered, spin-liquid-like regime at all temperatures with highly correlated magnetic moments, which are inequivalent on the two Dy sites in this material.
Direct comparison of the single crystal diffraction data (see Section~\ref{Sec_WISH}) and the scattering simulated for the different scattering planes is rather complicated as the Dy used in the WISH experiment is not isotopically enriched, and thus the effects of neutron absorption may play a large role.
Also, due to the limited $Q$ range of the data available for fitting, as well as the fact that it is unpolarised neutron data, the RMC methods used tend to overestimate the degree of ``diffuseness"~\cite{Paddison_2012,Paddison_2013}.

The radial spin correlation functions calculated through the RMC refinements are shown in Fig.~\ref{Fig4_correls} for the two different Dy sites in \sdo.
The data are obtained by averaging 10 independent spin configurations, although numerical variations between different configurations are rather small.
In the adopted system of labelling the set of the $i^\mathrm{th}$ neighbours with the index $i$ at a radial distance $r_i$, $r_1=3.43$~\AA\ refers to the nearest neighbour distance (legs of zigzag ladders), $r_2$ is the next-nearest neighbours distance (rungs of the ladders), $r_4$ is the shortest distance between the moments on different ladders, $r_5=2r_1$ is the next-nearest neighbours distance along the $c$~axis and so on.
Some of the interionic distances for the Dy2 site are labeled in Fig.~\ref{Fig1_structure}, right panel.
The correlation between the spins belonging to two different sites is identically zero because of the presumed strictly Ising character.
The data suggest that for the Dy2 site the nearest neighbour correlations are definitively \afm\ (automatically setting the $r_5$ correlations to be ferromagnetic), while the next-nearest correlations are significantly smaller but also \afm.
For the Dy1 site, the correlations are somewhat weaker, with the nearest-neighbour correlations being marginally ferromagnetic ($r_1$) while the next-nearest correlations ($r_2$) are stronger and \afm.
The inter-ladder correlation ($r_4$) are ferromagnetic for both sites and slightly more pronounced for the Dy1 site.
The average correlations beyond the distance of 20~\AA\ are statistically insignificant.
On warming to 0.4~K the spin-spin correlations remain very similar to what is calculated for $T=0.06$~K while with the temperature increasing to 0.8~K, they tend to marginally decrease.

\subsection{PND in an applied magnetic field}\label{In_field_PND}
\begin{figure}[tb] 
\centering
\includegraphics[width=\columnwidth]{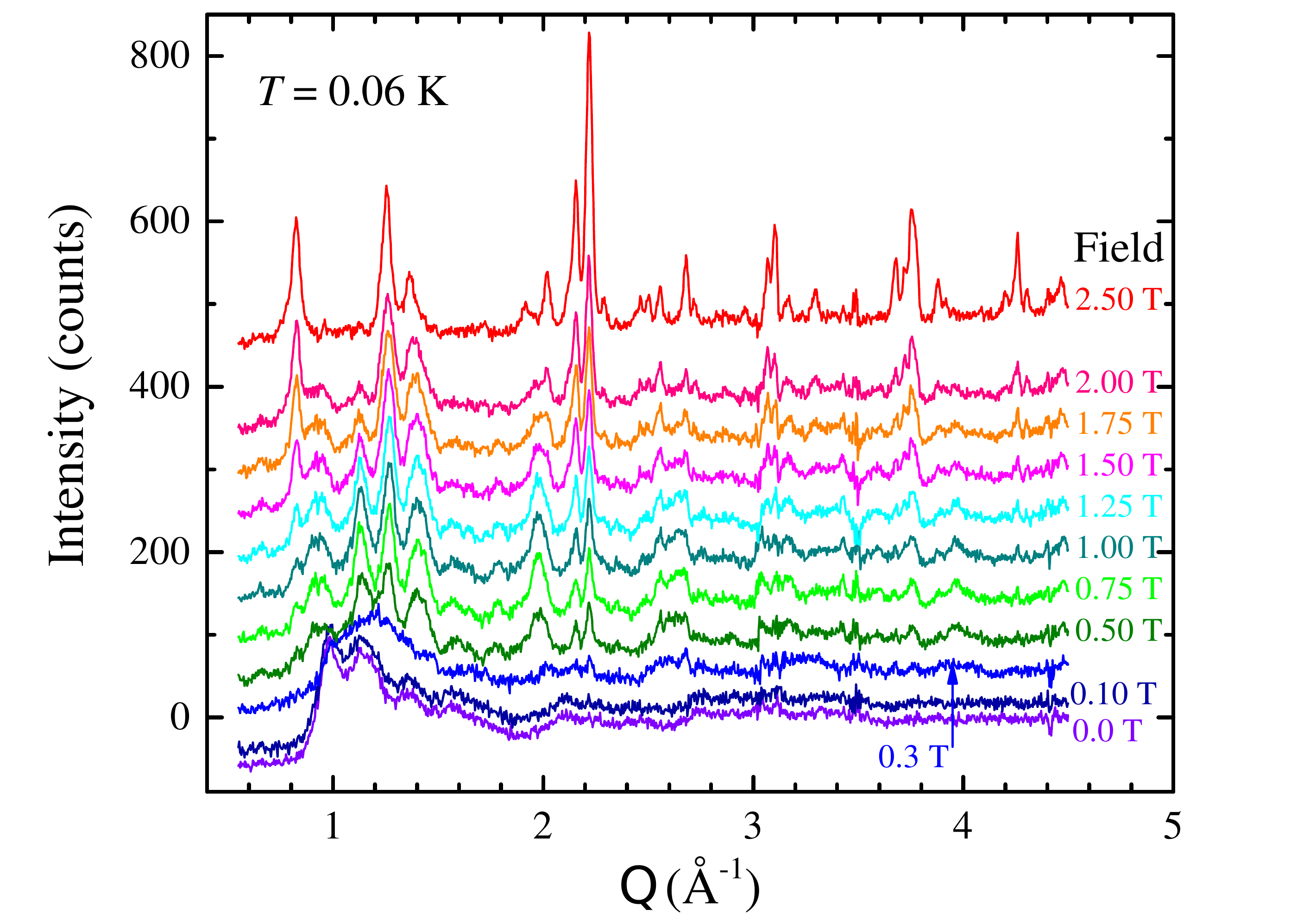}
\vspace{-5mm}
\caption{(Colour online) PND patterns of \sdo\ measured at $T=0.06$~K in different magnetic fields.
			The $T=10$~K background is subtracted from the data.
			The curves are consecutively offset by 100 counts per tesla for clarity.}
\label{Fig5_PND}
\end{figure}

Fig.~\ref{Fig5_PND} shows the evolution of the PND patterns of \sdo\ with an applied magnetic field at 0.06~K.
The data are obtained by subtracting the 10~K diffraction pattern in order to isolate the magnetic signal from the lattice contribution as well as the sample environment.
Given the discussion in the previous section, the 10~K data contain a considerable paramagnetic signal, which resulted in noticeable over-subtraction at low scattering angles. 

From Fig.~\ref{Fig5_PND} it follows that in lower applied fields (up to 0.3~T) the arrangement of the magnetic moments in \sdo\ are similar to what is observed in zero field; there are only very short-range correlations between the moments.
New, significantly sharper scattering features appear in fields around 0.5~T, while in fields above 2.0~T the scattering pattern consists only of resolution limited peaks. 
Thus the observed magnetic peaks can be separated into two different types through their field dependencies.
The first type are the peaks which start to appear with field increasing above 0.5~T and the intensity of which then starts to decrease in fields above 1~T.
These peaks are absent in the pattern measured at 2.5~T.
The intensity of the second type of the peaks continues to increase with increasing field.
From the positions of the Bragg peaks of this second type a magnetic propagation vector ${\bf k} = 0$ can be determined.
The first type of magnetic peaks originate from an \afm\ component to the magnetic order which is only stable at intermediate values of the magnetic field and which is more pronounced at 0.8~K than at 0.06~K.
Although in the refinements these highly structured \afm\ features are treated as regular Bragg peaks, their width is not resolution limited, therefore they do not imply the presence of a long range magnetic order.
The single crystal diffraction data shown below (see section~\ref{Sec_WISH}) provide more in-depth information on the magnetic correlations in the intermediate fields. 

\begin{figure}[tb] 
\centering
\vspace{-1 mm}
\includegraphics[width=\columnwidth]{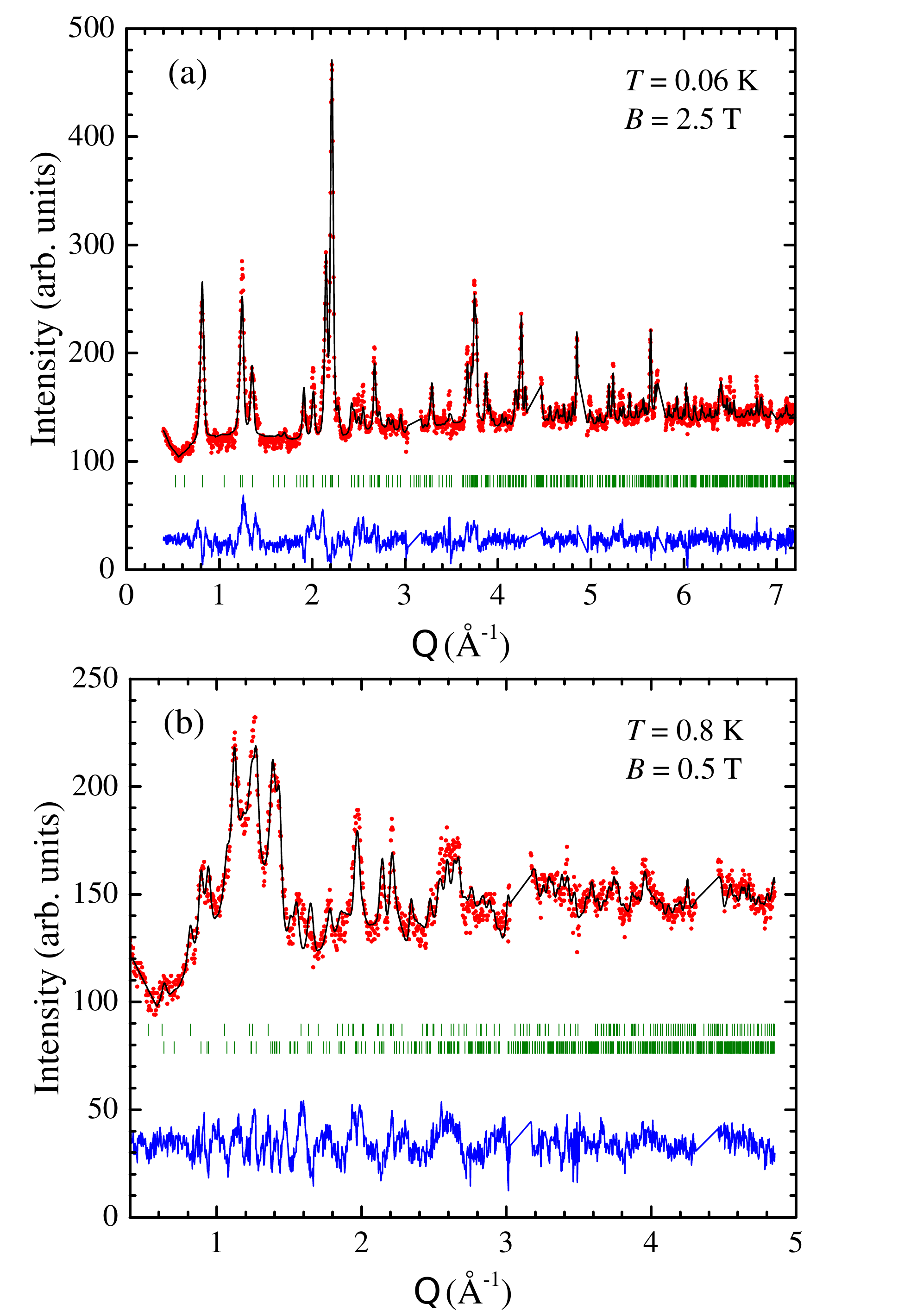}
\vspace{-5mm}
\caption{(Colour online) Rietveld refinement of the PND data in an applied magnetic field (a) at 0.06~K and 2.5~T, (b) at 0.8~K and 0.5~T.
The experimental data (red dots), calculated patterns (black lines) and the difference curves (blue lines) are shown.
Ticks represent the positions of the magnetic Bragg peaks.}
\label{Fig6_Rietveld}
\end{figure}

The magnetic propagation vector \KP\ was determined from the position of 9 magnetic reflections in a difference pattern created by subtracting the zero field data at 10~K from the 0.8~K data at 1.5~T.
As there are no relics of the magnetic peaks created through the magnetic propagation vector ${\bf k}^\prime$ in the spectrum recorded at 0.06~K and 2.5~T the corresponding data can be treated assuming a single magnetic phase.
For ${\bf k} = 0$, magnetic symmetry analysis was used to determine the irreducible representations (IR) and their basisvectors (BV) for the two Dy sites.
Different possibilities were tested against the data showing that a ferromagnetic structure with moments aligned along the $b$~axis gives the best fit to the measured intensities.
From the single crystal magnetisation measurements~\cite{Hayes_2012}, however, it is known that at 2.5~T and 0.5~K the values of magnetisation are $M\approx 2.0 \mu_{\rm B}$ for $B \parallel a$, $M\approx 6.0 \mu_{\rm B}$ for $B \parallel b$, and $M\approx 2.5 \mu_{\rm B}$ for $B \parallel c$.
It is therefore very likely that for a powder sample, after averaging over all directions of an applied field, only the strongest component of magnetisation ($B \parallel b$) becomes dominant, while all other magnetisation components become indistinguishable from the background.
In fact, a model with ferromagnetic structures on two different Dy sites pointing along $b$ and $c$ axes returned a fit to the measured intensities only marginally worse than the $M \parallel b$ model.

A refinement of the 0.06~K and 2.5~T data was done using the atomic coordinates determined from the high resolution D1A data for the nuclear phase and taking into account the extra reflections coming from the copper contained in the dilution refrigerator insert through a LeBail fit.
This refinement allowed us to extract a scale factor which was then used for refinements of the difference data sets containing solely the magnetic scattering contribution.
Fig.~\ref{Fig6_Rietveld}a displays the corresponding fit to the difference spectrum, dataset at 0.06~K and 2.5~T minus dataset at 10~K in zero field.
Magnetic moment values of 6.0(1) and 1.3(1)$\mu_{\rm B}$ were determined for the two Dy sites.

This result could be compared with previous observations, e.g., the \seo\ or \sho\ compounds~\cite{Petrenko_2008,Hayes_2011,Young_2012} where it was shown that only one out of the two Er or Ho sites possesses a sizeable magnetic moment (the PND data is insufficient to determine which of the two sites has the higher magnetic moment).
However, the comparison is not straightforward, as we are comparing the in-field behaviour of \sdo\ with zero-field properties of \seo\ and \sho.
The in-field behaviour of the latter two compounds is yet to be understood.
\begin{table}[tb] 
\caption{Irreducible representations (IR) and their basisvectors (BV) for \KP\ for the Wykoff position 4c of the space group $Pnam$.
	R (I) stands for real (imaginary) component of the BV.}
\begin{ruledtabular}
\begin{tabular}{lcccc}
\vspace{1mm}
\KP\		&	& BV1	& BV2	& BV3	\\	\hline
IR1		&	&		&		&		\\	\hline
$x, y, z$	&R	& 100	& 010	& 001	\\ 
		&I	& 000	& 000	& 000	\\	\hline
\vspace{1mm}
$-x+\frac{1}{2}, y+\frac{1}{2}, z+\frac{1}{2}$	&R	& -$\frac{1}{2}00$	& $0\frac{1}{2}0$	& $00\frac{1}{2}$	\\ 
\vspace{1mm}
		&I	& -$\frac{\sqrt{3}}{2}00$	& 0$\frac{\sqrt{3}}{2}0$	& 00$\frac{\sqrt{3}}{2}$	\\	\hline
IR2		&	&		&		&		\\	\hline
$x, y, z$	&R	& 100	& 010	& 001	\\ 
		&I	& 000	& 000	& 000	\\	\hline
\vspace{1mm}
$-x+\frac{1}{2}, y+\frac{1}{2}, z+\frac{1}{2}$	&R	& $\frac{1}{2}00$	& 0-$\frac{1}{2}0$	& 00-$\frac{1}{2}$	\\ 
\vspace{1mm}
		&I	& $\frac{\sqrt{3}}{2}00$	& 0-$\frac{\sqrt{3}}{2}0$	& 00-$\frac{\sqrt{3}}{2}$	\\		
\end{tabular}
\end{ruledtabular}
\label{TableI}
\end{table} 

As indicated above, the application of a magnetic field leads to the appearance of magnetic reflections originating from two different magnetic propagation vectors.
Within the range of temperature and field values covered in our measurements, there is no scattering pattern containing only the \afm\ contribution with \KP; there is always a contribution from \KO.
Both contributions have to be taken into account in a refinement.

Magnetic symmetry analysis within spacegroup $Pnam$ for \KO\ shows that the two Dy sites are each split into two independent orbits each having the same two allowed IRs.
As every IR is composed of three BVs and a phase factor between the orbits has to be refined, a total of 14 parameters are in theory available to describe the magnetic structure of this antiferromagnetic component.
This number of variables is too large in view of the limited number of magnetic reflections and therefore the magnetic moment values on the orbits originating from the same Dy site were constrained to be equal.
The difference data set (data at $T=0.8$~K, $B=0.5$~T minus 10~K zero-field background) was refined using two magnetic phases representing the \KO\ and \KP\ contributions and fixing the scale factors to the value previously determined from the refinement of the $T=0.06$~K, $B=2.5$~T data.
Testing the two IRs listed in Table~\ref{TableI} only a model using IR2 for the \KP\ contribution gives a good refinement.
As in the case of the \KO\ contribution only one of the two Dy sites has a significant magnetic moment which is predominantly pointing in the direction of the $b$~axis.

It is again not possible to determine from the PND data which of the two Dy sites carries the large magnetic moment.
Crystal field calculations~\cite{Poole_2014} on \sdo\ proposed the existence of a strong Ising anisotropy with the magnetic moments on the Dy1 site predominantly lying along the $c$~axis and those of the Dy2 site in the direction of the $b$~axis.
As our refinement shows the moment to be aligned along the direction of the $b$~axis, we assume the Dy2 site carries the large magnetic moment.
The phase between the two moments on the two orbits of the Dy2 site refines to $\phi/2\pi=0.34(1)$, Table~\ref{TableII} lists the refined coefficients of the BVs of the \KP\ contribution and the ferromagnetic component related to \KO, Fig.~\ref{Fig6_Rietveld}b displays a plot of the refinement.

\begin{table}[tb] 
\caption{Results of the two-phase refinements of the difference data sets for 0.8~K/0.5~T and 0.8~K/1.5~T with 10~K/0~T subtracted.}
\begin{ruledtabular}
\begin{tabular}{lc|lccc}
\vspace{1mm}
\KO\		& Ferro $\parallel b$	&  \KP\						& BV1	& BV2	& BV3	\\	\hline
\multicolumn{6}{l}{0.8 K/0.5 T} 														\\	\hline
Dy1		& -0.5(1)			&  Dy1						& --		& 0.6(2)	& -0.6(4)	\\
		& 				& \multicolumn{4}{l}{$\phi /2 \pi \; ({\rm Dy}1_1-{\rm Dy}1_2) =$-0.30(9)}	 \\	\hline		
Dy2		& -2.1(1)			&  Dy2						& --		& 5.8(1)	& -1.4(4)	\\
		& 				& \multicolumn{4}{l}{$\phi /2 \pi \; ({\rm Dy2_1 - Dy}2_2) =$-0.34(9)}	 \\	\hline		
\multicolumn{2}{l|}{$R_{\rm Mag}=24$\%}	& \multicolumn{4}{l}{$R_{\rm Mag}=20$\%}	 \\	\hline
\multicolumn{6}{l}{0.8 K/1.5 T} 														\\	\hline
Dy1		& -0.7(1)			&  Dy1						& --		& 0.8(1)	& --		\\
		& 				& \multicolumn{4}{l}{$\phi /2 \pi \; ({\rm Dy1_1 - Dy1_2}) =$-0.35(5)}	 \\	\hline		
Dy2		& -3.8(1)			&  Dy2						& --		& 8.2(1)	& --		\\
		& 				& \multicolumn{4}{l}{$\phi /2 \pi \; ({\rm Dy2_1- Dy}2_2) =$-0.333(5)}	 \\	\hline		
\multicolumn{2}{l|}{$R_{\rm Mag}=12$\%}	& \multicolumn{4}{l}{$R_{\rm Mag}=14$\%}	 \\	
\end{tabular}
\end{ruledtabular}
\label{TableII}
\end{table} 

\begin{figure}[b] 
\centering
\vspace{-10 mm}
\includegraphics[width=\columnwidth]{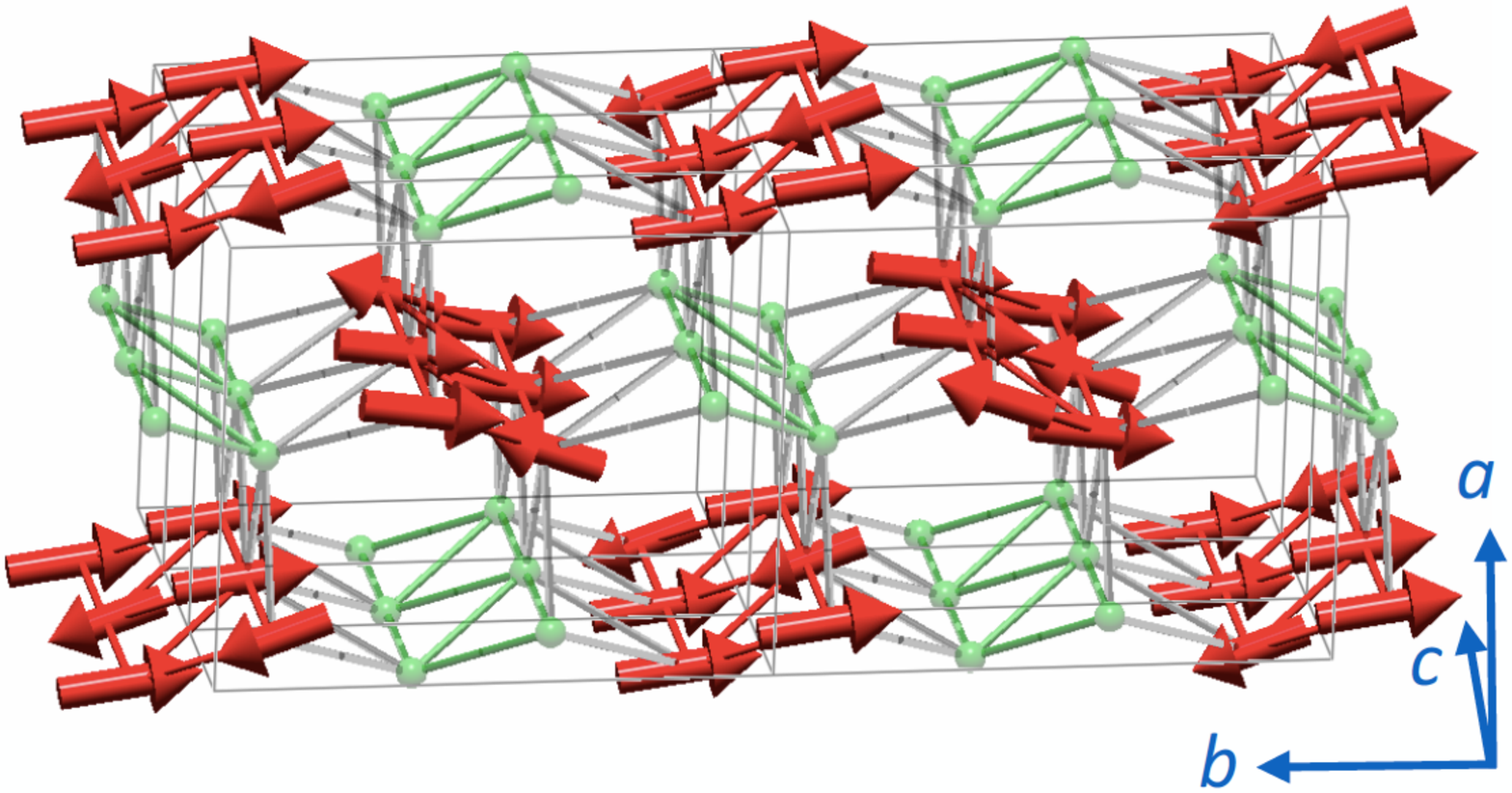}
\vspace{-10 mm}
\caption{(Colour online) Magnetic structure of \sdo\ at 0.8~K in an applied magnetic field of 0.5~T as obtained through the Rietveld refinement of the PND data.}
\label{Fig7_structure}
\end{figure}
Taking the \KP\ contribution on its own the magnetic structure would resemble an {\it up-up-down} configuration of the magnetic moments on the two legs of the honeycomb layers along the $c$~direction of the lattice with the magnetic moment value of the down position being twice as large as that of the up position.
The refined value of the phase $\phi /2 \pi \approx 0.333$ between the Dy2$_1$ and the Dy2$_2$ sites which each define one leg of the ladders (Fig.~\ref{Fig7_structure}) shows that this {\it up-up-down} configuration is in phase between the two legs of the zigzag ladder and therefore on the honeycomb lattice as a whole.
The magnetic structure resulting from the superposition of this sine-wave modulated \KP\ contribution and the ferromagnetic \KO\ contribution does not change this {\it up-up-down} configuration but now the magnetic moment values depending strongly on the size of the ferromagnetic contribution.
At 0.8~K and 0.5~T, the relative sizes of the \KO\ and \KP\ contributions result in nearly equal moment values along the chains of about 3.9 and -5.0$\mu_{\rm B}$.
Figure~\ref{Fig7_structure} displays the magnetic structure where the small values of the magnetic moments for the Dy1 site were set to zero in order to highlight the main features of the structure.
A refinement of the data taken at 0.8~K with a higher magnetic field of 1.5~T resulted in the values listed in Table~\ref{TableII}.
The {\it up-up-down} configuration is retained, however, the magnetic moment values (4.4 and -7.9$\mu_{\rm B}$) now differ considerably as the ferromagnetic contribution becomes more important.

In general, there is not enough justification for the introduction of an amplitude modulated magnetic structure in \sdo.
These structures are common in the RE metallic compounds (\sdo\ is an insulator) or at intermediate temperatures, where thermal fluctuations are important (while the proposed structure is for temperatures as low as 0.8 and 0.06~K).
The implication here is that the proposed structure returns the best fit to the observed PND patterns, but it is not necessarily the true magnetic structure of \sdo\ in an applied field.
Although the magnetic moments on the identical Dy sites are most likely to be equal in size in the actual structure, the main complication in the PND analysis is the highly anisotropic response of the system to an applied magnetic field.
This response is more adequately tested with the single crystal diffraction techniques. 

\section{Single crystal neutron diffraction}
\label{Sec_WISH}
\subsection{Zero field results}
\begin{figure}[tb] 
\centering
\vspace{-1 mm}
\includegraphics[width=1.02\columnwidth]{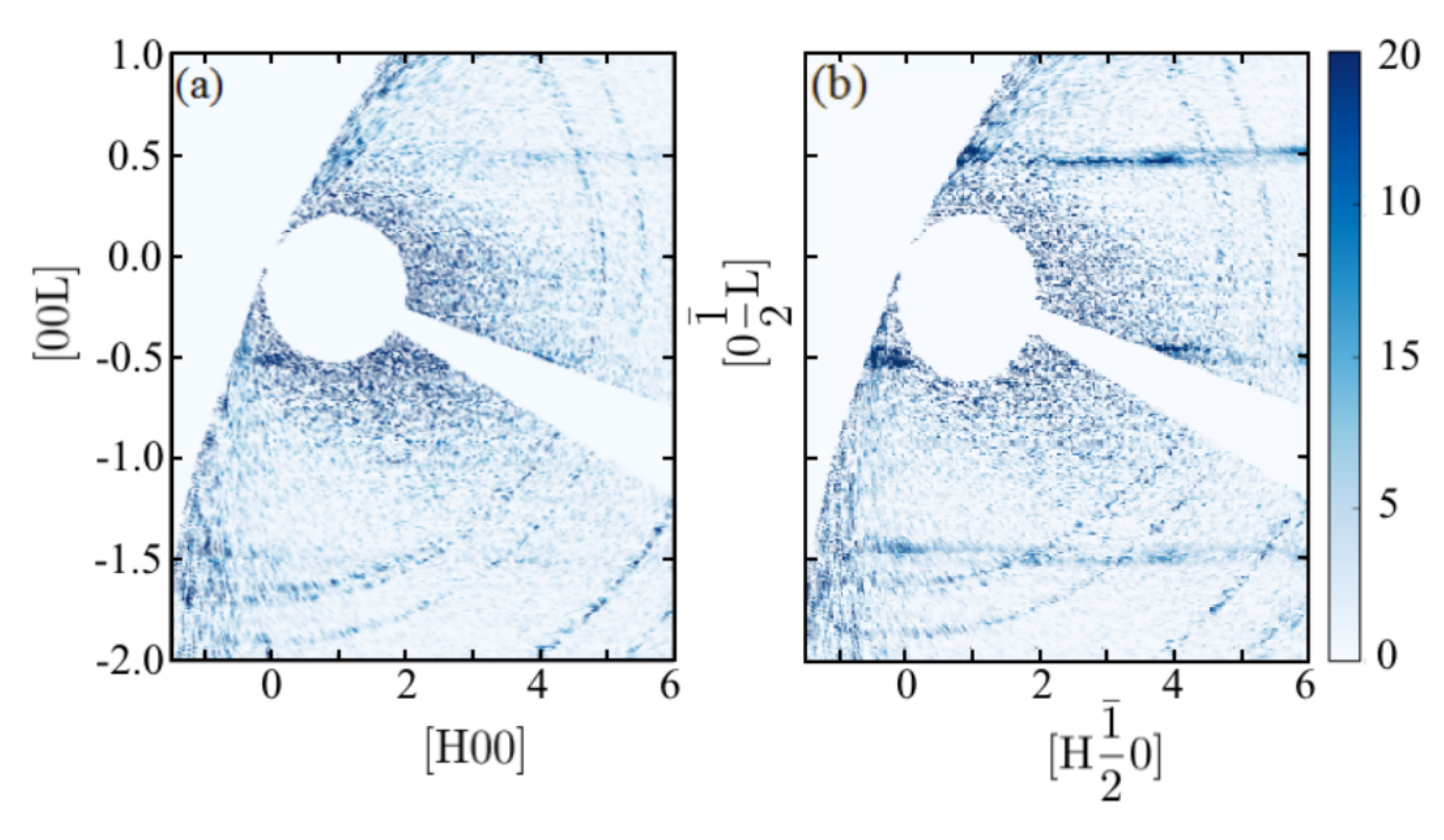}
\vspace{-8 mm}
\caption{(Colour online) Single crystal neutron diffraction maps of \sdo\ at $T=0.06$~K in the $(hkl)$ scattering plane for (a) $k=0.0 \pm 0.15$, and (b) $k=-0.5 \pm 0.15$.
		The magnetic component of the scattering is isolated by subtracting the 10~K background.
		As the subtraction procedure is not perfect, remnant powder lines from the copper sample holder can also be seen in all diffraction maps.}
\label{Fig8_ZeroField}
\end{figure}

Fig.~\ref{Fig8_ZeroField} shows the zero-field intensity maps of single crystal diffraction at base temperature, $T=0.06$~K.
Similar to what has been observed in \seo~\cite{Hayes_2011} and \sho~\cite{Young_2013}, in reciprocal space there are two-dimensional (2D) planes of scattering intensity, which after projection onto a particular scattering plane form ``rods" at $(h\,k \pm \! \frac{1}{2})$, $(h\,k \pm \! \frac{3}{2})$ and similar positions.
These 2D scattering features correspond to the formation of spin chains running along the $c$~axis with well-developed intrachain \afm\ correlations and almost no correlations between the chains.
From the very limited width of the ``rods" along the $[001]$ direction, approaching the resolution limit of the diffractometer, it follows that the \afm\ correlations along the chains extend over many unit cells.

The rods in \sdo\ are almost perfectly flat and show only very small modulations in their position and intensity as a function of $h$, but there is a noticeable variation of intensity as a function of $k$.
Broad intensity maxima are observed around the $k=\pm 0.5$ while for $k=0$ and $k=\pm 1$ the rods are barely visible.
This tendency -- the signal is more intense outside of the horizontal scattering plane -- is also preserved in an applied magnetic field (see next section) and was not observed in other compounds in the \sRo\ family where the scattering rods were clearly visible in the $(h0l)$ plane for \sho\ and the $(0kl)$ for \seo.
In both \seo\ and \sho, magnetic ions on only one crystallographic site, RE2, participate in the formation of the \afm\ chains (or ladders), while the moments on another site, RE1, tend to form the \KO\ \afm\ structures.
Interestingly, the RMC calculations, section~\ref{Zero_field_PND}, also indicate a small increase in intensity of the ``rods" outside of the $k=0$ plane, as can be seen in Fig.~\ref{Fig3_RMC}, bottom panel.

Additional measurements of diffraction intensity taken at 1.7~K have shown that the magnetic diffuse scattering signal in zero field has largely disappeared at this temperature.

\subsection{In-field results}
\begin{figure}[tb] 
\centering
\vspace{-1 mm}
\includegraphics[width=0.97\columnwidth]{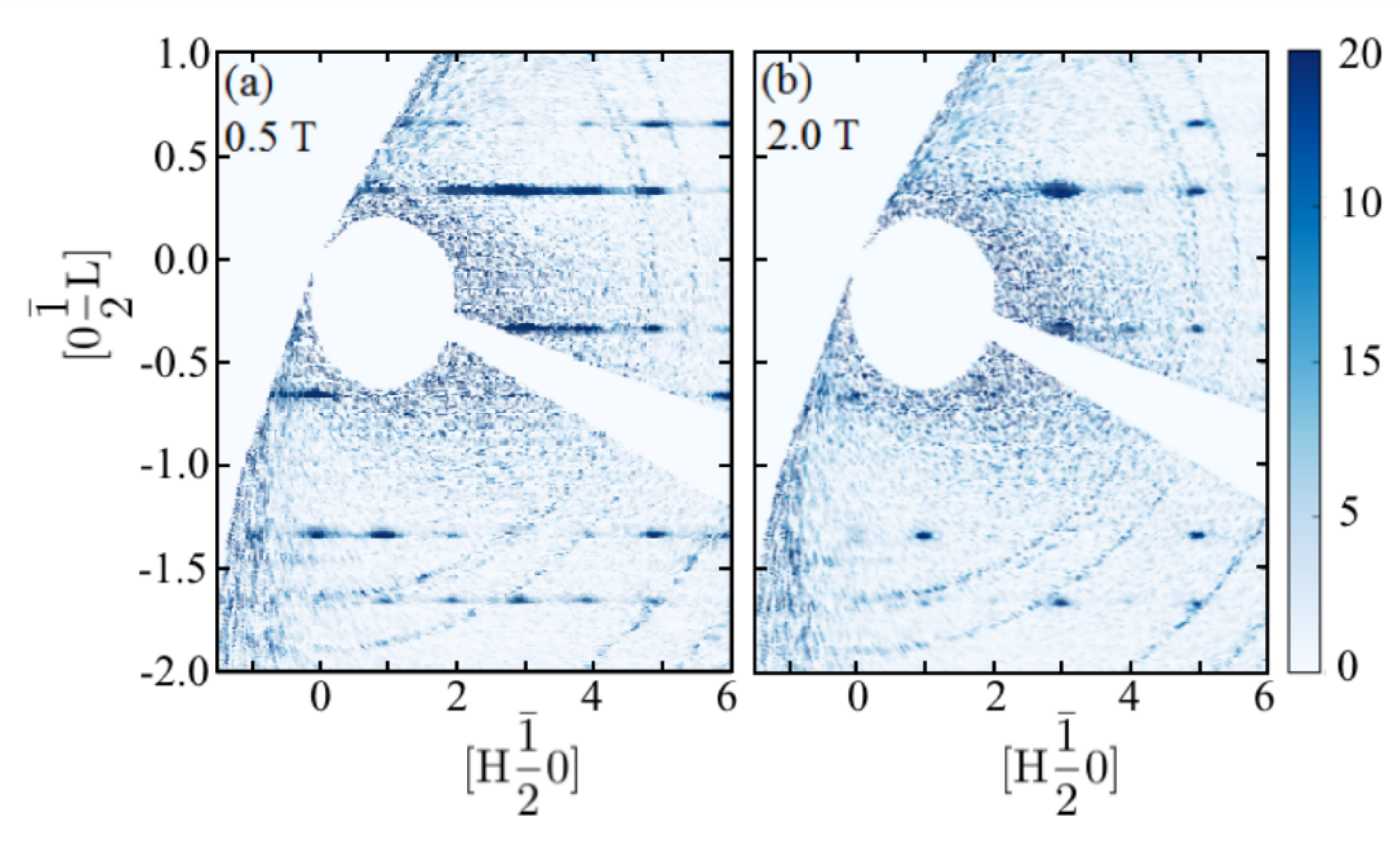}
\vspace{-3 mm}
\caption{(Colour online) Single crystal neutron diffraction intensity maps for \sdo\ at $T=0.06$~K in an applied field of (left panel) 0.5 and (right panel) 2.0~T.
		The maps were produced by integrating the scattering signal in the $(hkl)$ planes over the range of $k$ between $-0.65$ and $-0.35$, making them directly comparable to Fig.~\ref{Fig8_ZeroField}b for which $k=-0.5 \pm 0.15$.}
\label{Fig9_Field}
\end{figure}

Given that the diffuse signal is more pronounced outside of the $(h0l)$ scattering plane, Fig.~\ref{Fig9_Field} traces the field dependence of the magnetic intensity in the $(hkl)$ plane for $-0.65<k<-0.35$.
At $T=60$~mK, the measurements were taken in a field from zero to 3~T with a step-size of 0.5~T.
The data taken at 0.5, 1.0 and 1.5~T look almost identical, they are represented by the left-side panel in Fig.~\ref{Fig9_Field}.
In these fields, the rods move to the $(h \, k \, \frac{n}{3})$ positions, where $n=1, 2, 4, 5, 7...$ and gain considerable intensity modulation along the $h$ direction.
The maxima in intensity correspond to integer values of $h$, but the width of the diffuse scattering features along this direction remains considerably larger than the instruments resolution.
From fitting the diffuse scattering features to overlapping Gaussian peaks we estimate that the magnetic correlation length along the $h$ direction as approximately 7~\AA~\cite{width_remark}.
There is only a marginal increase (not more than 5\%) in the correlation length along the $h$ direction for the field increasing from 0.5 to 1.5~T. 

\begin{figure}[tb] 
\centering
\vspace{-5mm}
\includegraphics[width=\columnwidth]{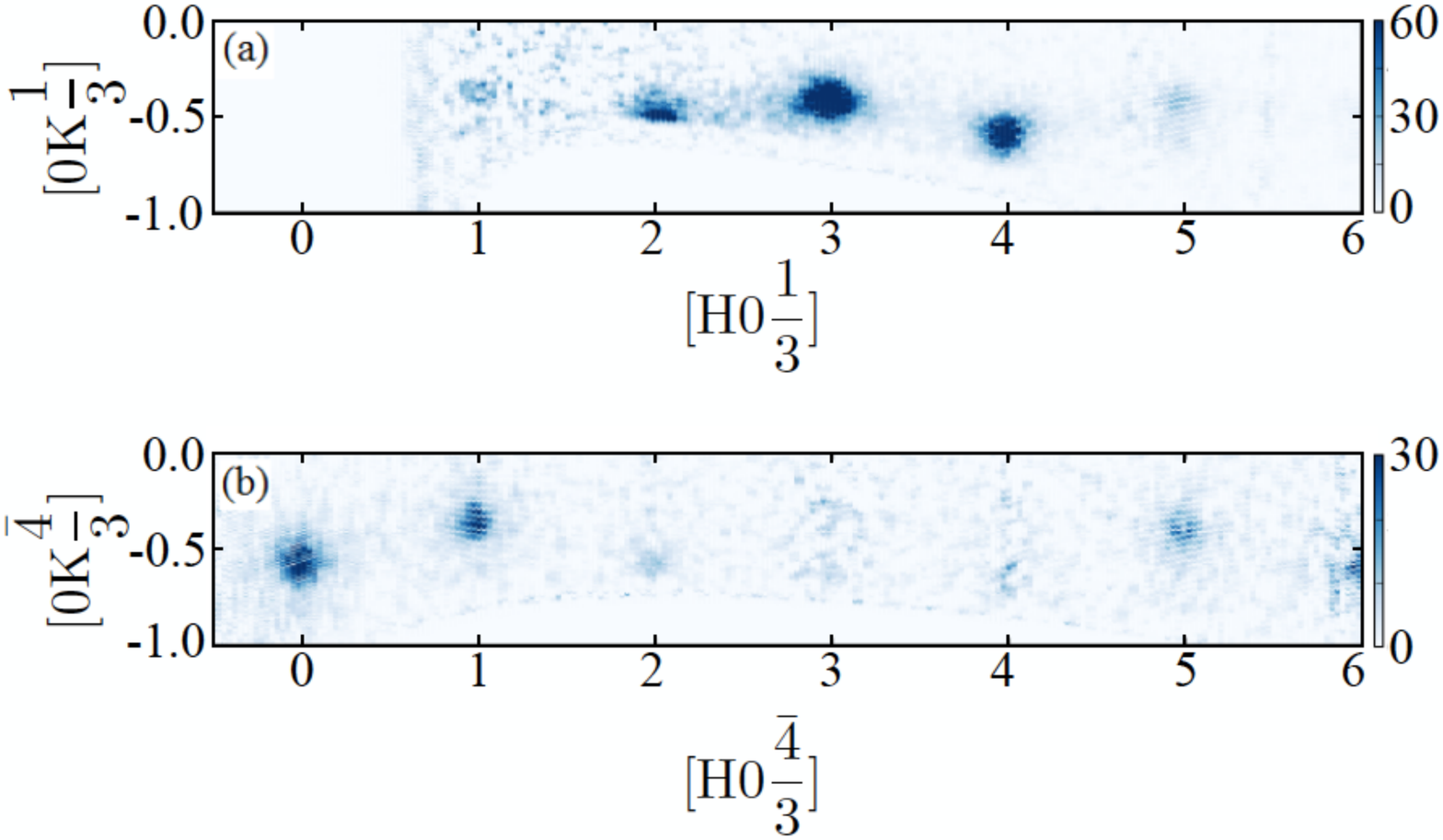}
\vspace{-10mm}
\caption{(Colour online) Single crystal neutron diffraction intensity maps for \sdo\ at $T=0.06$~K in an applied field of 1.0~T for (a) $(hk\frac{1}{3})$ and (b) $(hk\frac{\bar{4}}{3})$ scattering planes.}
\label{Fig10_hkl}
\end{figure}

Given the substantial coverage of the out-of-plane scattering available on the WISH diffractometer, we were also able to locate more accurately the positions of the intensity maxima in the $k$ direction.
Fig.~\ref{Fig10_hkl} shows the intensity maps for the $(hkl)$ planes for $l=\frac{1}{3}, \frac{\bar{4}}{3}$ and demostrates that in the intermediate-field regime the scattering remains diffuse (there are no properly formed Bragg peaks observed), but becomes much more localised for the $k$ as well as the $h$ directions.
The magnetic correlation length is similar for these two directions.
There is a pronounced undulation in the positions of the intensity maxima away from the zero-field $k=-\frac{1}{2}$ line, however, due to a finite width it is difficult to distinguish between the incommensurate positions of the type $k=-\frac{1}{2} \pm \delta$, with $\delta \approx 0.15$ and the $k=-\frac{1}{3}, -\frac{2}{3}$ predicted by the magnetic propagation vector \KP\ (see section~\ref{In_field_PND}).

A field 2.0~T marks the transition into a more polarised phase, and the pattern collected in this field (right-hand panel in Fig.~\ref{Fig9_Field}) contains almost resolution-limited features.
Above this field, in 2.5 and 3.0~T, the diffuse scattering signal completely disappears from the $(h \, \frac{1}{2} \, l)$ scattering plane.
In these higher fields, no diffuse scattering signal was detected anywhere in reciprocal space accessible on the WISH diffractometer.
Instead, the higher field data contain well-defined resolution-limited Bragg peaks in the $(h0l)$ plane with $h=$ even integer, $l=$ any integer, as well as in the $(h1l)$ and $(h\bar{1}l)$ planes with $h$ and $l$ being any integer numbers.

Additional measurements performed at 1.7~K showed that the intensities of the high-field magnetic Bragg peaks at this temperature are practically identical to those measured at base temperature, 0.06~K.
This observation is in agreement with the $H-T$ phase diagram proposed from the ultrasound measurements~\cite{Bidaud_2016}, where the critical field for the transition to the high-field phase is temperature independent in the range 0.05 to 1.8~K.
Presumably much higher temperatures are required to induce significant thermal disorder in the high-field magnetically-polarised phase.

The meaning of the observed scattering patterns is qualitatively rather transparent.
On application of moderate fields along the $b$ axis, the zero-field collection of uncorrelated \afm\ chains in \sdo\ is replaced by ferromagnetic {\it up-up-down} chains, which have significantly more developed inter-chain correlations.  
This intermediate-field regime corresponds to the development of a magnetisation plateau extending from about 0.2 to 2.0~T~\cite{Hayes_2012} where the arrangement of the magnetic moments remain practically constant.
Despite an increased correlation length (7~\AA\ is long enough to link the Dy2 ions in the different ladders) full long-range magnetic order is still missing in this regime. 

\begin{figure}[tb] 
\centering
\includegraphics[width=\columnwidth]{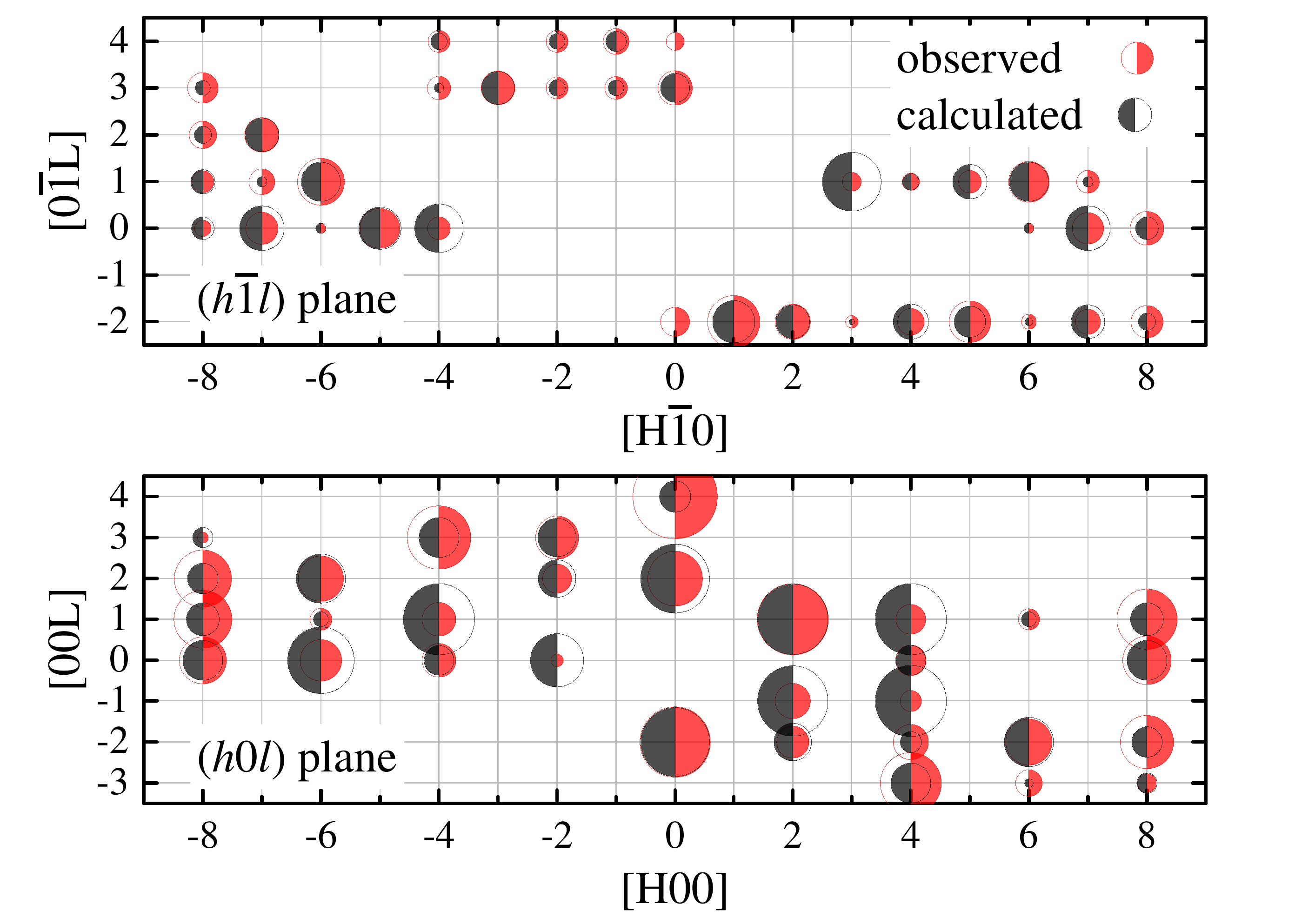}
\vspace{-5mm}
\caption{(Colour online) Graphical representation of the magnetic intensity in the $(h\bar 1 l)$ (top panel) and $(h0l)$ (bottom panel) scattering planes in a field of 3~T.
		The area of the circles represents the observed (red) and calculated (black) intensity of the magnetic Bragg peaks.
		The magnetic intensity is obtained by subtracting the high-temperature background from the $T=0.06$~K data.
		The calculated intensity is normalised to the magnetic form-factor for the Dy$^{3+}$ ions.}
\label{Fig11}
\end{figure}

The appearance of magnetic Bragg peaks at integer positions in higher fields can be explained by the field-induced ferromagnetic arrangement of the magnetic moments on the Dy2 sites along the field direction.
For the $(h1l)$ plane, the intensity of all nuclear peaks due to Dy atoms is extremely small because of a destructive interference between the Dy1 and Dy2 sites.
The intensity becomes significant as soon as the scattering amplitude from the two sites is allowed to differ.
One therefore has to conclude that only the Dy2 sites become fully ordered in fields above 2~T while the Dy1 sites remain largely disordered -- with the reported excitation gap of 4~meV~\cite{Poole_2014} for the Dy1 sites a much stronger field is required to induced a notable moment on them along the $b$ axis.

A full refinement of the magnetic structure in high fields is challenging, as the number of the observed reflections is relatively small and the effects of the wavelength-dependent absorption and extinction are difficult to quantify.
We therefore limit ourselves to a qualitative comparison of the observed Bragg peaks intensity to the intensity calculated for a simple model, in which the magnetic moments on the Dy2 sites are presumed to be fully polarised along the field while the ordered moment on the Dy1 sites is exactly zero.
The intensity is calculated presuming unit-length moments on the Dy2 sites, normalised to the magnetic form-factor for the Dy$^{3+}$ ions and then scaled up to match the measured intensity.
Fig.~\ref{Fig11} illustrates reasonable agreement between the experimental data and the calculations for such a model. 
The most noticeable disagreement in the $(h1l)$ plane is perhaps in the intensity of the $(012)$ and $(014)$ peaks, which are observed experimentally but absent in the calculations, while in the $(h0l)$ plane the intensities of the observed peaks at ($\pm$4 0 $\pm$1) are systematically lower than the calculated ones.
Allowing for partial polarisation of the Dy1 site (small moments all pointing either along the field or along the $c$~axis) does not improve the agreement.
In fact, in order to produce non-zero intensity for the $(012)$ and $(014)$ peaks one has to split the Dy1 and/or Dy2 sites and presume different magnetic moments on Dy1$_1$:Dy1$_2$ and/or Dy2$_1$:Dy2$_2$ sites.
This kind of splitting, however, also generates non-zero intensity for the $(h0l)$ peaks with odd values of $h$, which are systematically absent in the WISH data.
More accurate data collection using a single-wavelength diffractometer might be required in order to fully refine the field-induced structure of \sdo.

\section{Summary}
We have probed the development of the low-temperature magnetic correlations in \sdo\ using neutron diffraction techniques.

In zero field, \sdo\ demonstrates only short-range magnetic order down to the lowest experimentally available temperature, 0.06~K.
The RMC analysis of the neutron diffraction patterns observed on a powder sample reveals significant differences in the spin-spin correlations between the Dy$^{3+}$ ions belonging to two crystallographically inequivalent sites, similarly to what has previously been observed in other members of the \sRo\ family.
The formation of one-dimensional spin chains (or ladders) running along the $c$~axis on the Dy2 site has also been observed in a single crystal neutron diffraction experiment.

Application of an external magnetic field at low temperatures initially results in increased correlations in the short-range order regime and then in the appearance of Bragg-like features in the diffraction patterns corresponding to the formation of a longer-range magnetic order. 
Despite the natural limitations of the magnetic structure determination from PND in an applied magnetic field, Rietveld refinement gave clear indications as to the nature of the field-induced structures.
The evolution from a short-range order to a nearly ordered {\it up-up-down} configuration and then to a field-polarised state is most clearly seen in the single crystal neutron diffraction experiment for $H \parallel b$ direction.

\section*{ACKNOWLEDGMENTS}
We are grateful to B.Z.~Malkin for numerous discussions of the magnetic properties of \sdo\ and related compounds.
We would also like to acknowledge J.A.M.~Paddison for his help with RMC analysis as well as T.J.~Hayes for his help with sample preparations and neutron scattering measurements.
\bibliography{SrLn2O4_all}

\begin{thebibliography}{26}%
\makeatletter
\providecommand \@ifxundefined [1]{%
 \@ifx{#1\undefined}
}%
\providecommand \@ifnum [1]{%
 \ifnum #1\expandafter \@firstoftwo
 \else \expandafter \@secondoftwo
 \fi
}%
\providecommand \@ifx [1]{%
 \ifx #1\expandafter \@firstoftwo
 \else \expandafter \@secondoftwo
 \fi
}%
\providecommand \natexlab [1]{#1}%
\providecommand \enquote  [1]{``#1''}%
\providecommand \bibnamefont  [1]{#1}%
\providecommand \bibfnamefont [1]{#1}%
\providecommand \citenamefont [1]{#1}%
\providecommand \href@noop [0]{\@secondoftwo}%
\providecommand \href [0]{\begingroup \@sanitize@url \@href}%
\providecommand \@href[1]{\@@startlink{#1}\@@href}%
\providecommand \@@href[1]{\endgroup#1\@@endlink}%
\providecommand \@sanitize@url [0]{\catcode `\\12\catcode `\$12\catcode
  `\&12\catcode `\#12\catcode `\^12\catcode `\_12\catcode `\%12\relax}%
\providecommand \@@startlink[1]{}%
\providecommand \@@endlink[0]{}%
\providecommand \url  [0]{\begingroup\@sanitize@url \@url }%
\providecommand \@url [1]{\endgroup\@href {#1}{\urlprefix }}%
\providecommand \urlprefix  [0]{URL }%
\providecommand \Eprint [0]{\href }%
\providecommand \doibase [0]{http://dx.doi.org/}%
\providecommand \selectlanguage [0]{\@gobble}%
\providecommand \bibinfo  [0]{\@secondoftwo}%
\providecommand \bibfield  [0]{\@secondoftwo}%
\providecommand \translation [1]{[#1]}%
\providecommand \BibitemOpen [0]{}%
\providecommand \bibitemStop [0]{}%
\providecommand \bibitemNoStop [0]{.\EOS\space}%
\providecommand \EOS [0]{\spacefactor3000\relax}%
\providecommand \BibitemShut  [1]{\csname bibitem#1\endcsname}%
\let\auto@bib@innerbib\@empty
\bibitem [{\citenamefont {Diep}(2005)}]{Diep_2005}%
  \BibitemOpen
  \bibinfo {editor} {\bibfnamefont {H.~T.}\ \bibnamefont {Diep}},\ ed.,\
  \href@noop {} {\emph {\bibinfo {title} {Frustrated Spin Systems}}}\ (\bibinfo
   {publisher} {World Scientific, Singapore},\ \bibinfo {year}
  {2005})\BibitemShut {NoStop}%
\bibitem [{\citenamefont {Ramirez}(2001)}]{Buschow_2001}%
  \BibitemOpen
  \bibfield  {author} {\bibinfo {author} {\bibfnamefont {A.~P.}\ \bibnamefont
  {Ramirez}},\ }\href@noop {} {\emph {\bibinfo {title} {Physics of Magnetism
  and Magnetic Materials}}},\ edited by\ \bibinfo {editor} {\bibfnamefont
  {K.~H.~L.}\ \bibnamefont {Buschow}},\ Vol.~\bibinfo {volume} {13}\ (\bibinfo
  {publisher} {Elsevier, Amsterdam},\ \bibinfo {year} {2001})\ Chap.~\bibinfo
  {chapter} {4}\BibitemShut {NoStop}%
\bibitem [{\citenamefont {Lacroix}\ \emph {et~al.}(2011)\citenamefont
  {Lacroix}, \citenamefont {Mendels},\ and\ \citenamefont
  {Mila}}]{Lacroix_2011}%
  \BibitemOpen
  \bibinfo {editor} {\bibfnamefont {C.}~\bibnamefont {Lacroix}}, \bibinfo
  {editor} {\bibfnamefont {P.}~\bibnamefont {Mendels}}, \ and\ \bibinfo
  {editor} {\bibfnamefont {F.}~\bibnamefont {Mila}},\ eds.,\ \href@noop {}
  {\emph {\bibinfo {title} {Introduction to Frustrated Magnetism}}}\ (\bibinfo
  {publisher} {Springer-Verlag, Berlin},\ \bibinfo {year} {2011})\BibitemShut
  {NoStop}%
\bibitem [{\citenamefont {Ramirez}(1994)}]{Ramirez_1994}%
  \BibitemOpen
  \bibfield  {author} {\bibinfo {author} {\bibfnamefont {A.~P.}\ \bibnamefont
  {Ramirez}},\ }\href@noop {} {\bibfield  {journal} {\bibinfo  {journal} {Annu.
  Rev. Mater. Sci.}\ }\textbf {\bibinfo {volume} {24}},\ \bibinfo {pages} {453}
  (\bibinfo {year} {1994})}\BibitemShut {NoStop}%
\bibitem [{\citenamefont {Greedan}(2001)}]{Greedan_2001}%
  \BibitemOpen
  \bibfield  {author} {\bibinfo {author} {\bibfnamefont {J.~E.}\ \bibnamefont
  {Greedan}},\ }\href@noop {} {\bibfield  {journal} {\bibinfo  {journal} {J.
  Mater. Chem.}\ }\textbf {\bibinfo {volume} {11}},\ \bibinfo {pages} {37}
  (\bibinfo {year} {2001})}\BibitemShut {NoStop}%
\bibitem [{\citenamefont {Karunadasa}\ \emph {et~al.}(2005)\citenamefont
  {Karunadasa}, \citenamefont {Huang}, \citenamefont {Ueland}, \citenamefont
  {Lynn}, \citenamefont {Schiffer}, \citenamefont {Regan},\ and\ \citenamefont
  {Cava}}]{Karunadasa_2005}%
  \BibitemOpen
  \bibfield  {author} {\bibinfo {author} {\bibfnamefont {H.}~\bibnamefont
  {Karunadasa}}, \bibinfo {author} {\bibfnamefont {Q.}~\bibnamefont {Huang}},
  \bibinfo {author} {\bibfnamefont {B.~G.}\ \bibnamefont {Ueland}}, \bibinfo
  {author} {\bibfnamefont {J.~W.}\ \bibnamefont {Lynn}}, \bibinfo {author}
  {\bibfnamefont {P.}~\bibnamefont {Schiffer}}, \bibinfo {author}
  {\bibfnamefont {K.~A.}\ \bibnamefont {Regan}}, \ and\ \bibinfo {author}
  {\bibfnamefont {R.~J.}\ \bibnamefont {Cava}},\ }\href@noop {} {\bibfield
  {journal} {\bibinfo  {journal} {Phys. Rev. B}\ }\textbf {\bibinfo {volume}
  {71}},\ \bibinfo {pages} {144414} (\bibinfo {year} {2005})}\BibitemShut
  {NoStop}%
\bibitem [{\citenamefont {Petrenko}(2014)}]{Petrenko_2014}%
  \BibitemOpen
  \bibfield  {author} {\bibinfo {author} {\bibfnamefont {O.~A.}\ \bibnamefont
  {Petrenko}},\ }\href@noop {} {\bibfield  {journal} {\bibinfo  {journal} {Low
  Temp. Phys.}\ }\textbf {\bibinfo {volume} {40}},\ \bibinfo {pages} {106}
  (\bibinfo {year} {2014})}\BibitemShut {NoStop}%
\bibitem [{\citenamefont {Petrenko}\ \emph {et~al.}(2008)\citenamefont
  {Petrenko}, \citenamefont {Balakrishnan}, \citenamefont {Wilson},
  \citenamefont {de~Brion}, \citenamefont {Suard},\ and\ \citenamefont
  {Chapon}}]{Petrenko_2008}%
  \BibitemOpen
  \bibfield  {author} {\bibinfo {author} {\bibfnamefont {O.~A.}\ \bibnamefont
  {Petrenko}}, \bibinfo {author} {\bibfnamefont {G.}~\bibnamefont
  {Balakrishnan}}, \bibinfo {author} {\bibfnamefont {N.~R.}\ \bibnamefont
  {Wilson}}, \bibinfo {author} {\bibfnamefont {S.}~\bibnamefont {de~Brion}},
  \bibinfo {author} {\bibfnamefont {E.}~\bibnamefont {Suard}}, \ and\ \bibinfo
  {author} {\bibfnamefont {L.~C.}\ \bibnamefont {Chapon}},\ }\href@noop {}
  {\bibfield  {journal} {\bibinfo  {journal} {Phys. Rev. B}\ }\textbf {\bibinfo
  {volume} {78}},\ \bibinfo {pages} {184410} (\bibinfo {year}
  {2008})}\BibitemShut {NoStop}%
\bibitem [{\citenamefont {Hayes}\ \emph {et~al.}(2011)\citenamefont {Hayes},
  \citenamefont {Balakrishnan}, \citenamefont {Deen}, \citenamefont {Manuel},
  \citenamefont {Chapon},\ and\ \citenamefont {Petrenko}}]{Hayes_2011}%
  \BibitemOpen
  \bibfield  {author} {\bibinfo {author} {\bibfnamefont {T.~J.}\ \bibnamefont
  {Hayes}}, \bibinfo {author} {\bibfnamefont {G.}~\bibnamefont {Balakrishnan}},
  \bibinfo {author} {\bibfnamefont {P.~P.}\ \bibnamefont {Deen}}, \bibinfo
  {author} {\bibfnamefont {P.}~\bibnamefont {Manuel}}, \bibinfo {author}
  {\bibfnamefont {L.~C.}\ \bibnamefont {Chapon}}, \ and\ \bibinfo {author}
  {\bibfnamefont {O.~A.}\ \bibnamefont {Petrenko}},\ }\href@noop {} {\bibfield
  {journal} {\bibinfo  {journal} {Phys. Rev. B}\ }\textbf {\bibinfo {volume}
  {84}},\ \bibinfo {pages} {174435} (\bibinfo {year} {2011})}\BibitemShut
  {NoStop}%
\bibitem [{\citenamefont {Young}\ \emph {et~al.}(2012)\citenamefont {Young},
  \citenamefont {Chapon},\ and\ \citenamefont {Petrenko}}]{Young_2012}%
  \BibitemOpen
  \bibfield  {author} {\bibinfo {author} {\bibfnamefont {O.}~\bibnamefont
  {Young}}, \bibinfo {author} {\bibfnamefont {L.~C.}\ \bibnamefont {Chapon}}, \
  and\ \bibinfo {author} {\bibfnamefont {O.~A.}\ \bibnamefont {Petrenko}},\
  }\href@noop {} {\bibfield  {journal} {\bibinfo  {journal} {J. Phys.: Conf.
  Ser.}\ }\textbf {\bibinfo {volume} {391}},\ \bibinfo {pages} {012081}
  (\bibinfo {year} {2012})}\BibitemShut {NoStop}%
\bibitem [{\citenamefont {Young}\ \emph {et~al.}(2013)\citenamefont {Young},
  \citenamefont {Wildes}, \citenamefont {Manuel}, \citenamefont {Ouladdiaf},
  \citenamefont {Khalyavin}, \citenamefont {Balakrishnan},\ and\ \citenamefont
  {Petrenko}}]{Young_2013}%
  \BibitemOpen
  \bibfield  {author} {\bibinfo {author} {\bibfnamefont {O.}~\bibnamefont
  {Young}}, \bibinfo {author} {\bibfnamefont {A.~R.}\ \bibnamefont {Wildes}},
  \bibinfo {author} {\bibfnamefont {P.}~\bibnamefont {Manuel}}, \bibinfo
  {author} {\bibfnamefont {B.}~\bibnamefont {Ouladdiaf}}, \bibinfo {author}
  {\bibfnamefont {D.~D.}\ \bibnamefont {Khalyavin}}, \bibinfo {author}
  {\bibfnamefont {G.}~\bibnamefont {Balakrishnan}}, \ and\ \bibinfo {author}
  {\bibfnamefont {O.~A.}\ \bibnamefont {Petrenko}},\ }\href {\doibase
  10.1103/PhysRevB.88.024411} {\bibfield  {journal} {\bibinfo  {journal} {Phys.
  Rev. B}\ }\textbf {\bibinfo {volume} {88}},\ \bibinfo {pages} {024411}
  (\bibinfo {year} {2013})}\BibitemShut {NoStop}%
\bibitem [{\citenamefont {Quintero-Castro}\ \emph {et~al.}(2012)\citenamefont
  {Quintero-Castro}, \citenamefont {Lake}, \citenamefont {Reehuis},
  \citenamefont {Niazi}, \citenamefont {Ryll}, \citenamefont {Islam},
  \citenamefont {Fennell}, \citenamefont {Kimber}, \citenamefont {Klemke},
  \citenamefont {Ollivier}, \citenamefont {Sakai}, \citenamefont {Deen},\ and\
  \citenamefont {Mutka}}]{Quintero_2012}%
  \BibitemOpen
  \bibfield  {author} {\bibinfo {author} {\bibfnamefont {D.~L.}\ \bibnamefont
  {Quintero-Castro}}, \bibinfo {author} {\bibfnamefont {B.}~\bibnamefont
  {Lake}}, \bibinfo {author} {\bibfnamefont {M.}~\bibnamefont {Reehuis}},
  \bibinfo {author} {\bibfnamefont {A.}~\bibnamefont {Niazi}}, \bibinfo
  {author} {\bibfnamefont {H.}~\bibnamefont {Ryll}}, \bibinfo {author}
  {\bibfnamefont {A.~T. M.~N.}\ \bibnamefont {Islam}}, \bibinfo {author}
  {\bibfnamefont {T.}~\bibnamefont {Fennell}}, \bibinfo {author} {\bibfnamefont
  {S.~A.~J.}\ \bibnamefont {Kimber}}, \bibinfo {author} {\bibfnamefont
  {B.}~\bibnamefont {Klemke}}, \bibinfo {author} {\bibfnamefont
  {J.}~\bibnamefont {Ollivier}}, \bibinfo {author} {\bibfnamefont {V.~G.}\
  \bibnamefont {Sakai}}, \bibinfo {author} {\bibfnamefont {P.~P.}\ \bibnamefont
  {Deen}}, \ and\ \bibinfo {author} {\bibfnamefont {H.}~\bibnamefont {Mutka}},\
  }\href@noop {} {\bibfield  {journal} {\bibinfo  {journal} {Phys. Rev. B}\
  }\textbf {\bibinfo {volume} {86}},\ \bibinfo {pages} {064203} (\bibinfo
  {year} {2012})}\BibitemShut {NoStop}%
\bibitem [{\citenamefont {Young}\ \emph {et~al.}(2014)\citenamefont {Young},
  \citenamefont {Balakrishnan}, \citenamefont {Lees},\ and\ \citenamefont
  {Petrenko}}]{Young_2014}%
  \BibitemOpen
  \bibfield  {author} {\bibinfo {author} {\bibfnamefont {O.}~\bibnamefont
  {Young}}, \bibinfo {author} {\bibfnamefont {G.}~\bibnamefont {Balakrishnan}},
  \bibinfo {author} {\bibfnamefont {M.~R.}\ \bibnamefont {Lees}}, \ and\
  \bibinfo {author} {\bibfnamefont {O.~A.}\ \bibnamefont {Petrenko}},\ }\href
  {\doibase 10.1103/PhysRevB.90.094421} {\bibfield  {journal} {\bibinfo
  {journal} {Phys. Rev. B}\ }\textbf {\bibinfo {volume} {90}},\ \bibinfo
  {pages} {094421} (\bibinfo {year} {2014})}\BibitemShut {NoStop}%
\bibitem [{\citenamefont {Cheffings}\ \emph {et~al.}(2013)\citenamefont
  {Cheffings}, \citenamefont {Lees}, \citenamefont {Balakrishnan},\ and\
  \citenamefont {Petrenko}}]{Cheffings_2013}%
  \BibitemOpen
  \bibfield  {author} {\bibinfo {author} {\bibfnamefont {T.~H.}\ \bibnamefont
  {Cheffings}}, \bibinfo {author} {\bibfnamefont {M.~R.}\ \bibnamefont {Lees}},
  \bibinfo {author} {\bibfnamefont {G.}~\bibnamefont {Balakrishnan}}, \ and\
  \bibinfo {author} {\bibfnamefont {O.~A.}\ \bibnamefont {Petrenko}},\
  }\href@noop {} {\bibfield  {journal} {\bibinfo  {journal} {J. Phys.: Condens.
  Matter}\ }\textbf {\bibinfo {volume} {25}},\ \bibinfo {pages} {256001}
  (\bibinfo {year} {2013})}\BibitemShut {NoStop}%
\bibitem [{\citenamefont {Fennell}\ \emph {et~al.}(2014)\citenamefont
  {Fennell}, \citenamefont {Pomjakushin}, \citenamefont {Uldry}, \citenamefont
  {Delley}, \citenamefont {Pr{\'e}vost}, \citenamefont {D{\'e}silets-Benoit},
  \citenamefont {Bianchi}, \citenamefont {Bewley}, \citenamefont {Hansen},
  \citenamefont {Klimczuk}, \citenamefont {Cava},\ and\ \citenamefont
  {Kenzelmann}}]{Poole_2014}%
  \BibitemOpen
  \bibfield  {author} {\bibinfo {author} {\bibfnamefont {A.}~\bibnamefont
  {Fennell}}, \bibinfo {author} {\bibfnamefont {V.~Y.}\ \bibnamefont
  {Pomjakushin}}, \bibinfo {author} {\bibfnamefont {A.}~\bibnamefont {Uldry}},
  \bibinfo {author} {\bibfnamefont {B.}~\bibnamefont {Delley}}, \bibinfo
  {author} {\bibfnamefont {B.}~\bibnamefont {Pr{\'e}vost}}, \bibinfo {author}
  {\bibfnamefont {A.}~\bibnamefont {D{\'e}silets-Benoit}}, \bibinfo {author}
  {\bibfnamefont {A.~D.}\ \bibnamefont {Bianchi}}, \bibinfo {author}
  {\bibfnamefont {R.~I.}\ \bibnamefont {Bewley}}, \bibinfo {author}
  {\bibfnamefont {B.~R.}\ \bibnamefont {Hansen}}, \bibinfo {author}
  {\bibfnamefont {T.}~\bibnamefont {Klimczuk}}, \bibinfo {author}
  {\bibfnamefont {R.~J.}\ \bibnamefont {Cava}}, \ and\ \bibinfo {author}
  {\bibfnamefont {M.}~\bibnamefont {Kenzelmann}},\ }\href@noop {} {\bibfield
  {journal} {\bibinfo  {journal} {Phys. Rev. B}\ }\textbf {\bibinfo {volume}
  {89}},\ \bibinfo {pages} {224511} (\bibinfo {year} {2014})}\BibitemShut
  {NoStop}%
\bibitem [{\citenamefont {Hayes}\ \emph {et~al.}(2012)\citenamefont {Hayes},
  \citenamefont {Young}, \citenamefont {Balakrishnan},\ and\ \citenamefont
  {Petrenko}}]{Hayes_2012}%
  \BibitemOpen
  \bibfield  {author} {\bibinfo {author} {\bibfnamefont {T.~J.}\ \bibnamefont
  {Hayes}}, \bibinfo {author} {\bibfnamefont {O.}~\bibnamefont {Young}},
  \bibinfo {author} {\bibfnamefont {G.}~\bibnamefont {Balakrishnan}}, \ and\
  \bibinfo {author} {\bibfnamefont {O.~A.}\ \bibnamefont {Petrenko}},\
  }\href@noop {} {\bibfield  {journal} {\bibinfo  {journal} {J. Phys. Soc.
  Japan}\ }\textbf {\bibinfo {volume} {84}},\ \bibinfo {pages} {024708}
  (\bibinfo {year} {2012})}\BibitemShut {NoStop}%
\bibitem [{\citenamefont {Bidaud}\ \emph {et~al.}(2016)\citenamefont {Bidaud},
  \citenamefont {Simard}, \citenamefont {Quirion}, \citenamefont {Pr\'evost},
  \citenamefont {Daneau}, \citenamefont {Bianchi}, \citenamefont {Dabkowska},\
  and\ \citenamefont {Quilliam}}]{Bidaud_2016}%
  \BibitemOpen
  \bibfield  {author} {\bibinfo {author} {\bibfnamefont {C.}~\bibnamefont
  {Bidaud}}, \bibinfo {author} {\bibfnamefont {O.}~\bibnamefont {Simard}},
  \bibinfo {author} {\bibfnamefont {G.}~\bibnamefont {Quirion}}, \bibinfo
  {author} {\bibfnamefont {B.}~\bibnamefont {Pr\'evost}}, \bibinfo {author}
  {\bibfnamefont {S.}~\bibnamefont {Daneau}}, \bibinfo {author} {\bibfnamefont
  {A.~D.}\ \bibnamefont {Bianchi}}, \bibinfo {author} {\bibfnamefont {H.~A.}\
  \bibnamefont {Dabkowska}}, \ and\ \bibinfo {author} {\bibfnamefont {J.~A.}\
  \bibnamefont {Quilliam}},\ }\href {\doibase 10.1103/PhysRevB.93.060404}
  {\bibfield  {journal} {\bibinfo  {journal} {Phys. Rev. B}\ }\textbf {\bibinfo
  {volume} {93}},\ \bibinfo {pages} {060404} (\bibinfo {year}
  {2016})}\BibitemShut {NoStop}%
\bibitem [{\citenamefont {Paddison}\ \emph {et~al.}(2013)\citenamefont
  {Paddison}, \citenamefont {Stewart},\ and\ \citenamefont
  {Goodwin}}]{Paddison_2013}%
  \BibitemOpen
  \bibfield  {author} {\bibinfo {author} {\bibfnamefont {J.~A.~M.}\
  \bibnamefont {Paddison}}, \bibinfo {author} {\bibfnamefont {J.~R.}\
  \bibnamefont {Stewart}}, \ and\ \bibinfo {author} {\bibfnamefont {A.~L.}\
  \bibnamefont {Goodwin}},\ }\href
  {http://stacks.iop.org/0953-8984/25/i=45/a=454220} {\bibfield  {journal}
  {\bibinfo  {journal} {J. Phys.: Condens. Matter}\ }\textbf {\bibinfo {volume}
  {25}},\ \bibinfo {pages} {454220} (\bibinfo {year} {2013})}\BibitemShut
  {NoStop}%
\bibitem [{\citenamefont {Wen}\ \emph {et~al.}(2015)\citenamefont {Wen},
  \citenamefont {Tian}, \citenamefont {Garlea}, \citenamefont {Koohpayeh},
  \citenamefont {McQueen}, \citenamefont {Li}, \citenamefont {Yan},
  \citenamefont {Rodriguez-Rivera}, \citenamefont {Vaknin},\ and\ \citenamefont
  {Broholm}}]{Wen_2015}%
  \BibitemOpen
  \bibfield  {author} {\bibinfo {author} {\bibfnamefont {J.-J.}\ \bibnamefont
  {Wen}}, \bibinfo {author} {\bibfnamefont {W.}~\bibnamefont {Tian}}, \bibinfo
  {author} {\bibfnamefont {V.~O.}\ \bibnamefont {Garlea}}, \bibinfo {author}
  {\bibfnamefont {S.~M.}\ \bibnamefont {Koohpayeh}}, \bibinfo {author}
  {\bibfnamefont {T.~M.}\ \bibnamefont {McQueen}}, \bibinfo {author}
  {\bibfnamefont {H.-F.}\ \bibnamefont {Li}}, \bibinfo {author} {\bibfnamefont
  {J.-Q.}\ \bibnamefont {Yan}}, \bibinfo {author} {\bibfnamefont {J.~A.}\
  \bibnamefont {Rodriguez-Rivera}}, \bibinfo {author} {\bibfnamefont
  {D.}~\bibnamefont {Vaknin}}, \ and\ \bibinfo {author} {\bibfnamefont {C.~L.}\
  \bibnamefont {Broholm}},\ }\href {\doibase 10.1103/PhysRevB.91.054424}
  {\bibfield  {journal} {\bibinfo  {journal} {Phys. Rev. B}\ }\textbf {\bibinfo
  {volume} {91}},\ \bibinfo {pages} {054424} (\bibinfo {year}
  {2015})}\BibitemShut {NoStop}%
\bibitem [{\citenamefont {Malkin}\ \emph {et~al.}(2015)\citenamefont {Malkin},
  \citenamefont {Nikitin}, \citenamefont {Mumdzhi}, \citenamefont {Zverev},
  \citenamefont {Yusupov}, \citenamefont {Gilmutdinov}, \citenamefont
  {Batulin}, \citenamefont {Gabbasov}, \citenamefont {Kiiamov}, \citenamefont
  {Adroja}, \citenamefont {Young},\ and\ \citenamefont
  {Petrenko}}]{Malkin_2015}%
  \BibitemOpen
  \bibfield  {author} {\bibinfo {author} {\bibfnamefont {B.~Z.}\ \bibnamefont
  {Malkin}}, \bibinfo {author} {\bibfnamefont {S.~I.}\ \bibnamefont {Nikitin}},
  \bibinfo {author} {\bibfnamefont {I.~E.}\ \bibnamefont {Mumdzhi}}, \bibinfo
  {author} {\bibfnamefont {D.~G.}\ \bibnamefont {Zverev}}, \bibinfo {author}
  {\bibfnamefont {R.~V.}\ \bibnamefont {Yusupov}}, \bibinfo {author}
  {\bibfnamefont {I.~F.}\ \bibnamefont {Gilmutdinov}}, \bibinfo {author}
  {\bibfnamefont {R.}~\bibnamefont {Batulin}}, \bibinfo {author} {\bibfnamefont
  {B.~F.}\ \bibnamefont {Gabbasov}}, \bibinfo {author} {\bibfnamefont {A.~G.}\
  \bibnamefont {Kiiamov}}, \bibinfo {author} {\bibfnamefont {D.~T.}\
  \bibnamefont {Adroja}}, \bibinfo {author} {\bibfnamefont {O.}~\bibnamefont
  {Young}}, \ and\ \bibinfo {author} {\bibfnamefont {O.~A.}\ \bibnamefont
  {Petrenko}},\ }\href {\doibase 10.1103/PhysRevB.92.094415} {\bibfield
  {journal} {\bibinfo  {journal} {Phys. Rev. B}\ }\textbf {\bibinfo {volume}
  {92}},\ \bibinfo {pages} {094415} (\bibinfo {year} {2015})}\BibitemShut
  {NoStop}%
\bibitem [{\citenamefont {Dublenych}(2016)}]{Dublenych_2016}%
  \BibitemOpen
  \bibfield  {author} {\bibinfo {author} {\bibfnamefont {Y.~I.}\ \bibnamefont
  {Dublenych}},\ }\href {\doibase 10.1103/PhysRevB.93.054415} {\bibfield
  {journal} {\bibinfo  {journal} {Phys. Rev. B}\ }\textbf {\bibinfo {volume}
  {93}},\ \bibinfo {pages} {054415} (\bibinfo {year} {2016})}\BibitemShut
  {NoStop}%
\bibitem [{\citenamefont {Balakrishnan}\ \emph {et~al.}(2009)\citenamefont
  {Balakrishnan}, \citenamefont {Hayes}, \citenamefont {Petrenko},\ and\
  \citenamefont {$\rm M^cK$~Paul}}]{Balakrishnan_2009}%
  \BibitemOpen
  \bibfield  {author} {\bibinfo {author} {\bibfnamefont {G.}~\bibnamefont
  {Balakrishnan}}, \bibinfo {author} {\bibfnamefont {T.~J.}\ \bibnamefont
  {Hayes}}, \bibinfo {author} {\bibfnamefont {O.~A.}\ \bibnamefont {Petrenko}},
  \ and\ \bibinfo {author} {\bibfnamefont {D.}~\bibnamefont {$\rm
  M^cK$~Paul}},\ }\href@noop {} {\bibfield  {journal} {\bibinfo  {journal} {J.
  Phys.: Condens. Matter}\ }\textbf {\bibinfo {volume} {21}},\ \bibinfo {pages}
  {012202} (\bibinfo {year} {2009})}\BibitemShut {NoStop}%
\bibitem [{\citenamefont {Chapon}\ \emph {et~al.}(2011)\citenamefont {Chapon},
  \citenamefont {Manuel}, \citenamefont {Radaelli}, \citenamefont {Benson},
  \citenamefont {Perrott}, \citenamefont {Ansell}, \citenamefont {Rhodes},
  \citenamefont {Raspino}, \citenamefont {Duxbury}, \citenamefont {Spill},
  \citenamefont {J},\ and\ \citenamefont {Norris}}]{WISH}%
  \BibitemOpen
  \bibfield  {author} {\bibinfo {author} {\bibfnamefont {L.~C.}\ \bibnamefont
  {Chapon}}, \bibinfo {author} {\bibfnamefont {P.}~\bibnamefont {Manuel}},
  \bibinfo {author} {\bibfnamefont {P.~G.}\ \bibnamefont {Radaelli}}, \bibinfo
  {author} {\bibfnamefont {C.}~\bibnamefont {Benson}}, \bibinfo {author}
  {\bibfnamefont {L.}~\bibnamefont {Perrott}}, \bibinfo {author} {\bibfnamefont
  {S.}~\bibnamefont {Ansell}}, \bibinfo {author} {\bibfnamefont
  {N.}~\bibnamefont {Rhodes}}, \bibinfo {author} {\bibfnamefont
  {D.}~\bibnamefont {Raspino}}, \bibinfo {author} {\bibfnamefont
  {D.}~\bibnamefont {Duxbury}}, \bibinfo {author} {\bibfnamefont
  {E.}~\bibnamefont {Spill}}, \bibinfo {author} {\bibnamefont {J}}, \ and\
  \bibinfo {author} {\bibnamefont {Norris}},\ }\href@noop {} {\bibfield
  {journal} {\bibinfo  {journal} {Neutron News}\ }\textbf {\bibinfo {volume}
  {22}},\ \bibinfo {pages} {22} (\bibinfo {year} {2011})}\BibitemShut {NoStop}%
\bibitem [{\citenamefont {Rodríguez-Carvajal}(1993)}]{FULLPROF}%
  \BibitemOpen
  \bibfield  {author} {\bibinfo {author} {\bibfnamefont {J.}~\bibnamefont
  {Rodríguez-Carvajal}},\ }\href@noop {} {\bibfield  {journal} {\bibinfo
  {journal} {Physica B: Condensed Matter}\ }\textbf {\bibinfo {volume} {192}},\
  \bibinfo {pages} {55 } (\bibinfo {year} {1993})}\BibitemShut {NoStop}%
\bibitem [{\citenamefont {Paddison}\ and\ \citenamefont
  {Goodwin}(2012)}]{Paddison_2012}%
  \BibitemOpen
  \bibfield  {author} {\bibinfo {author} {\bibfnamefont {J.~A.~M.}\
  \bibnamefont {Paddison}}\ and\ \bibinfo {author} {\bibfnamefont {A.~L.}\
  \bibnamefont {Goodwin}},\ }\href {\doibase 10.1103/PhysRevLett.108.017204}
  {\bibfield  {journal} {\bibinfo  {journal} {Phys. Rev. Lett.}\ }\textbf
  {\bibinfo {volume} {108}},\ \bibinfo {pages} {017204} (\bibinfo {year}
  {2012})}\BibitemShut {NoStop}%
\bibitem [{wid()}]{width_remark}%
  \BibitemOpen
  \href@noop {} {}\bibinfo {note} {Given that the instrument's contribution to
  the peak broadening is significantly smaller than the width of the diffuse
  scattering features, the magnetic correlation length is calculated as a
  reciprocal of the measured width.}\BibitemShut {Stop}%
\end{thebibliography}%
\end{document}